\documentclass[preprint,5.4pt]{elsarticle}
\usepackage[paperwidth=19.2cm,paperheight=26.2cm,hmargin=1.35cm,vmargin=1.65cm]{geometry}
\usepackage{tcolorbox}
\usepackage{multicol}
\usepackage[T1]{fontenc} 
\usepackage{printlen}

\usepackage{amssymb}
\usepackage{amsmath}
\usepackage[breaklinks,colorlinks,citecolor=blue,linkcolor=teal]{hyperref}
\usepackage{upgreek}
\usepackage{bm}

\DeclareMathAlphabet{\mathbfsf}{\encodingdefault}{\sfdefault}{bx}{sl}
\newcommand{\tens}[1]{\mathbfsf{#1}}
\newcommand{\bcdot}{\boldsymbol{\cdot}}
\newcommand{\bnabla}{\boldsymbol{\nabla}}

\newcommand{\bnhat}{\mathbf{\hat{n}}}

\journal{Journal of Computational Physics}

\begin{document}

\begin{frontmatter}



\title{A front-tracking immersed-boundary framework for simulating Lagrangian melting problems}

\author[label1]{Kevin Zhong\corref{cor1}}
\ead{k.zhong@utwente.nl}

\author[label1,label2]{Christopher. J. Howland}

\author[label1,label3]{Detlef Lohse}

\author[label1,label4]{Roberto Verzicco}

\cortext[cor1]{Corresponding author:}

\affiliation[label1]{organization={Physics of Fluids, University of Twente},
            addressline={Drienerlolaan 5}, 
            city={Enschede},
            postcode={7522NB}, 
            country={The Netherlands}}

\affiliation[label2]{organization={School of Mathematics and Statistics, University College Dublin},
            addressline={Belfield}, 
            city={Dublin 4},
            country={Ireland}}

\affiliation[label3]{organization={Max Planck Institute for Dynamics and Self-Organization},
            addressline={Am Fassberg 17},
            city={G\"{o}ttingen},
            postcode={37077},
            country={Germany}}

\affiliation[label4]{organization={Dipartimento di Ingegneria Industriale, University of Roma `Tor Vergata'},
            addressline={Via del Politecnico 1},
            city={Roma},
            postcode={00133},
            country={Italy}}

\begin{abstract}
In so-called Lagrangian melting problems, a solid immersed in a fluid medium is free to rotate and translate in tandem with its phase-change from solid to liquid. Such configurations may be classified as a fluid-solid interaction (FSI) problem coupled to phase-change. Our present work proposes a numerical method capable of simulating these Lagrangian melting problems and adopts a front-tracking immersed-boundary (IB) method. We use the moving least squares IB framework, a well-established method for simulating a diverse range of FSI problems \citep{vanella2009,detullio2016} and extend this framework to accommodate melting by additionally imposing the Stefan condition at the interface. In the spirit of canonical front-tracking methods, the immersed solid is represented by a discrete triangulated mesh which is separate from the Eulerian mesh in which the governing flow equations are solved. A known requirement for these methods is the need for comparable Eulerian and Lagrangian grid spacings to stabilise interpolation and spreading operations between the two grids. For a melting object, this requirement is inevitably violated unless interventional remeshing is introduced. Our work therefore presents a novel dynamic remeshing procedure to overcome this. The remeshing is based on a gradual coarsening of the triangulated Lagrangian mesh and amounts to a negligible computational burden per timestep owing to the incremental and local nature of its operations, making it a scalable approach. Moreover, the coarsening is coupled to a volume-conserving smoothing procedure detailed by \citet{kuprat2001}, ensuring a zero net volume change in the remeshing step to machine precision. This added feature makes our present method highly specialised to the study of melting problems, where precise measurements of the melting solid's volume  is often the primary predictive quantity of interest.
\end{abstract}

\begin{graphicalabstract}
\begin{center}
	\includegraphics[width=0.9\textwidth]{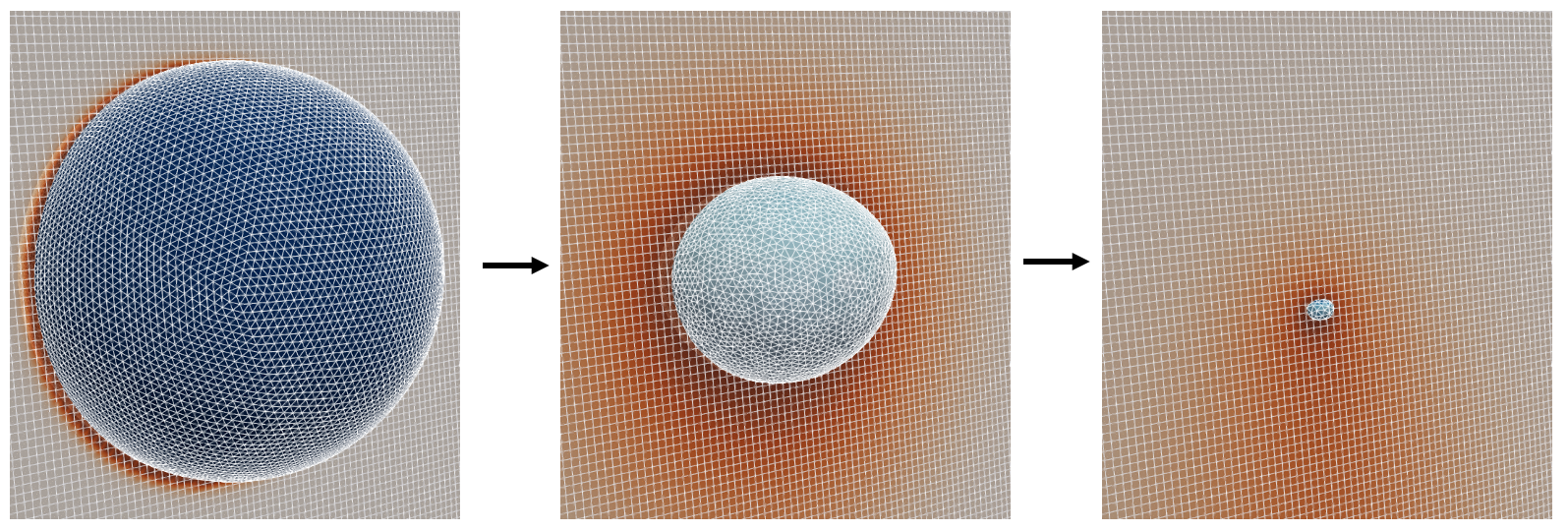}
\end{center}
\end{graphicalabstract}


\begin{keyword}
Immersed boundary method \sep Phase change \sep Fluid-solid interaction \sep Turbulent flows \sep Multiphase flow \sep One-fluid model \sep Eulerian--Lagrangian \sep Heat transfer \sep Meshing \sep Stefan problems



\end{keyword}

\end{frontmatter}


\section{Introduction}\label{sec:intro}
Many fluid flows encountered in nature and industry are coupled to a change of phase. Specifically in the case of melting, a phase-change is induced through a supply of sensible heat, typically from the liquid phase which overcomes the latent heat of fusion from the solid phase to cause the change in phase \citep{worster2000,davis2001,malyarenko2020}. The liquid medium will in general exert hydrodynamic forces and torques on the solid, causing the solid-body to rotate and translate, thereby influencing its melting dynamics. Such a flow configuration, which we call Lagrangian melting, may be classified as a fluid-solid interaction (FSI) problem coupled to phase-change. Here, we distinguish between fluid-solid interaction and fluid-structure interaction problems, where the latter usually entails solving complex dynamical equations which govern the structural deformation \citep{hou2012,griffith2020}. Fluid-solid interaction problems, meanwhile, treat the structure as a homogeneous, rigid solid with its motion governed by the Newton--Euler equations \citep{uhlmann2005} and is the treatment we consider for Lagrangian melting.

A prolific body of literature has developed for the simulation of phase-change problems. Recent years have favoured one-fluid formulations, where the physical system is described by a single set of equations encompassing both phases, and the interface boundary is represented by an appropriate distributed singularity in the governing equations \citep{prosperetti2009,tryggvason2011,mirjalili2017}. Broadly, one-fluid formulations can be categorised into interface-capturing or front-tracking (interface-tracking) schemes. Although both formulations have seen extensive application to phase-change problems \citep{juric1996,juric1998,chen1997,beckermann1999,udaykumar1999,al2002,boettinger2002,hester2020,huang2021,lyu2021}, when one considers FSI in tandem with phase-change, the extension of interface-capturing methods to accommodate FSI is non-trivial, although we note some recent works which have attempted this \citep{lecrivain2018,reder2021}. The key difficulty in applying interface-capturing schemes to FSI are attributed to the fact that the interface location is not explicitly tracked, but instead \textit{captured}, typically through an Eulerian phase-indicator field. This ambiguity in identifying the interface location poses a bottleneck if one wishes to solve FSI problems, where knowledge of the solid geometry and surface-integrated hydrodynamic loads are required to solve the dynamical equations governing the immersed solid's motion (i.e. Newton's laws of motion).

Front-tracking methods by contrast, do not suffer from the shortcomings of interface-capturing schemes in interface location ambiguity and therefore provide a viable avenue for simulating Lagrangian melting. In front-tracking, the solid-liquid interface location is explicitly represented by a Lagrangian mesh, separate from the Eulerian mesh in which the governing fluid flow equations are solved. Typically in 3D, this Lagrangian mesh takes the form of a triangulated surface \citep{juric1996,juric1998,tryggvason2001}, and is an unambiguous representation of the solid geometry, thereby providing a convenient framework for solving the equations governing the solid body's motion. Indeed, when the interface in question is a solid-liquid interface as with melting, front-tracking methods are essentially identical to immersed boundary methods (IBMs). IBMs enjoy widespread use owing to their versatility and relative simplicity in handling flows with complex geometries \citep{mittal2005,mittal2023,verzicco2023}. This has consequently led to a rich body of literature developing concerning the application of IBMs to flows involving FSI \citep{griffith2020}. 

This availability of well-established IBM techniques for simulating FSI problems will motivate us in the present work to propose a front-tracking immersed-boundary (IB) approach for simulating Lagrangian melting. Our present method will adopt the moving least squares (MLS) IB formulation, which has already been demonstrated as a versatile framework capable of handling a wide array of FSI problems \citep{vanella2009,detullio2016,wang2019,viola2020}. This method is of the direct-forcing category in IB methods \citep{fadlun2000} and the novelty of our present work will be extending this MLS-IB framework to accommodate phase-change (melting), thereby encompassing the full Lagrangian melting problem. At the core of direct-forcing IB methods are transfer operations between the Eulerian and Lagrangian mesh. 
This involves computing loads at discrete locations on the Lagrangian mesh, and appropriately spreading these loads to the neighbouring Eulerian cells \citep{mittal2023,verzicco2023}. To perform these operations stably, it is known that the typical spacing of these Lagrangian nodes (i.e. the Lagrangian mesh spacing) should be commensurate to the Eulerian grid spacing \citep{tryggvason2011,detullio2016,viola2020}. When the Lagrangian mesh in question represents a melting solid object, the Lagrangian mesh spacing inevitably shrinks and violates this condition unless interventional remeshing is applied. One approach is to adaptively refine the Eulerian mesh in regions local to the immersed solid as it melts. However, this abandons one of the primary appeals of canonical IB methods. That is, the use of structured Eulerian grids in which simple, efficient algorithms may be relied upon for solving the governing flow equations \citep{verzicco2023}.

Alternatively, one may keep the Eulerian grid fixed and instead apply remeshing operations to the Lagrangian mesh progressively as the object melts. This is the approach we adopt in our present work and is the crucial ingredient for successfully simulating Lagrangian melting problems. In particular, our remeshing procedure is based on local incremental coarsening of the Lagrangian mesh, followed by local smoothing to avoid numerically-undesirable features such as highly-skewed mesh elements or sharp geometric corners. Many surface remeshing algorithms perform remeshing through a global optimisation procedure, often reorganising both the connectivity and spatial coordinates of the mesh in a single step \citep{botsch2010}. Although this can tend to produce higher-quality meshes globally, this computation is more expensive and generally modifies the entirety of the mesh, even in locations where remeshing intervention may have been unnecessary. The latter consequence is particularly undesirable for simulations, as this effectively introduces artificial global deformations of the solid. By contrast, our present approach instead adopts an incremental local procedure to both coarsening and smoothing, only applying interventional remeshing to regions of the mesh where necessary. Consequently, the remeshing operations amount to an insignificant computational burden per timestep, making our method a highly scalable approach to larger problem sizes which is in the spirit of canonical IB methods. Moreover, our remeshing operations will adopt the volume-conserving smoothing algorithm of \citet{kuprat2001}, which, when coupled to mesh-coarsening can allow for one to achieve a net-zero volume change during remeshing to machine precision \citep{roghair2016}. This feature will be highly advantageous for the study of melting in particular, where the solid volume as a function of time is often the principal quantity of interest for prediction \citep{cenedese2023}. Since our remeshing procedure progressively coarsens the solid as it melts, a loss of geometric fidelity and hence, solution accuracy is inevitable as the solid melts down to scales approaching the Eulerian grid. We will demonstrate that this can be mitigated provided the simulations begin with a well-resolved Lagrangian mesh that is gradually coarsened, providing a good compromise between solution accuracy and computational efficiency.

Practical melting problems in natural and industrial environments are highly-complex physical systems involving many degrees of freedom and control parameters to fully account for \citep{malyarenko2020,cenedese2023}.
Inevitably, this complexity necessitates a restriction in the scope of what is to be covered by our present work.
Particular limitations are discussed in the following summary:
\begin{itemize}
    \item In modelling the motion of the solid objects, fluid and solid densities are assumed to be equal, such that buoyancy does not play a role.
    This restriction is primarily implemented due to our use of simple, periodic domains, where a buoyant solid object would rise perpetually.
    Adding a buoyancy force to the solid motion equations does not entail additional difficulties, but is beyond our current focus.
    \item We assume that thermal conductivity in the solid and liquid phases are constant and equal.
    For water, this is not strictly accurate, since the solid ice phase conducts heat four times faster than the liquid phase.
    However in practice, since the heat transport by the liquid is far more efficient than the conduction-limited transport by the solid, it is common to even fully neglect the heat flux through the ice when estimating ice-ocean melt rates \citep{dinniman2016}.
    The nature of the immersed boundary method also means that implementing different conductivities for the two phases is feasible \citep{verzicco2002}, but we leave this for future work.
    \item We restrict ourselves to the problem of a single solid object immersed in the fluid medium.
    This avoids the complication of collision modelling, which is required in the case of multiple immersed objects.
    Many studies have detailed how collisions can be addressed in interface-resolved IBM simulations \citep{kempe2012b,costa2015,ardekani2016}, and we foresee that coupling these to our melting interface would require relatively minor modifications.
    \item Changes in the interface topology are not considered.
    The current method therefore, cannot handle the potential break-up of a melting solid into multiple pieces.
    Handling topology changes is largely considered one of the greatest difficulties in front-tracking and is a difficulty we do not consider here \citep{tryggvason2011}.
    Typically, we shall restrict ourselves to simple, convex geometries in flow configurations where inhomogeneous melting effects are not too significant, such that topology changes are not expected.
\end{itemize}

This paper is organised as follows. In \S \ref{sec:goveqn}, we will outline the governing equations and associated boundary conditions to be solved. In \S \ref{sec:numerical} we provide details on the numerical methods adopted to solve these equations. We then detail our remeshing procedure in \S \ref{sec:remesh}, followed by an overall summary of all the steps in our numerical solver in \S \ref{sec:num_overview}. Test case results are presented in \S \ref{sec:tests}, followed by conclusions in \S \ref{sec:conclusions}.
\section{Governing equations}\label{sec:goveqn}
We consider an incompressible flow, alongside a transport equation for the temperature field $\theta$ using a 3D Cartesian coordinate system $\bm{x} = (x,y,z)$. In the fluid domain, the governing equations read:
\begin{align}\label{eq:goveqn1}
&\bnabla \bcdot \bm{u} = 0, \\
&\frac{\partial \bm{u}}{\partial t} + \bnabla \bcdot (\bm{u u} ) = -\frac{1}{\rho_f} \bnabla p + \nu \nabla^2 \bm{u} + \bm{f} + \bm{\mathcal{F}}, \label{eq:navstokes} \\
& \frac{\partial  \theta}{\partial t} + \bnabla \bcdot(\bm{u} \theta)  = \kappa_f  \nabla^2 \theta + f^\theta \label{eq:goveqn2}
\end{align}
where $\bm{u} = (u,v,w)$ is the velocity, $p$ is the pressure, $t$ is time, $\rho_f$ is the fluid density, $\nu$ is the fluid's kinematic viscosity, and $\kappa_f$ is the fluid's thermal diffusivity. Here, $\bm{f}$ and $f^\theta$ are body forces associated with the IB for the velocity field and temperature respectively and are prescribed to locally enforce the correct boundary conditions at the solid-liquid interface. The forcing $\bm{\mathcal{F}}$ in the momentum equation is an external body-force that may be prescribed in the system of interest. We will restrict our present work to the case where the fluid and solid phases have matched density and thermal diffusivity, i.e. $\rho_f = \rho_s$, $\kappa_f = \kappa_s$ and the subscripts will be omitted for brevity henceforth. Inside the solid phase, heat transfer is driven solely through conduction in the reference frame of the moving solid velocity, $\bm{U}_s$, such that inside the solid, its temperature distribution is governed by:
\begin{equation}\label{eq:goveqn_tsolid}
    \frac{\partial  \theta}{\partial t} + \bnabla \bcdot(\bm{U}_s \theta)  = \kappa  \nabla^2 \theta + f^\theta.
\end{equation}
The solid-body velocity has contributions from the linear translation of its centroid, $\bm{U}_c$, and rotation about its centroid which is characterised by its angular velocity $\bm{\Omega}_c$:
\begin{equation}\label{eq:Usolid}
    \bm{u}(\bm{x} = \bm{x}_s) \equiv \bm{U}_s = \bm{U}_c + \bm{\Omega}_c \times \bm{r}
\end{equation}
where $\bm{r} \equiv \bm{x}_s - \bm{X}_c$ is a vector representing the position of an arbitrary point inside the solid, $\bm{x}_s$, relative to the solid's centroid coordinate, $\bm{X}_c$. The solid centroid, $\bm{X}_c(t)$ will in general vary with time owing to both melting and the solid dynamical motion. Throughout this work, we will use upper-case letters to denote Lagrangian quantities, and lower-case for Eulerian field quantities. The solid object's translation and rotation are governed by the Newton--Euler equations:
\begin{equation}\label{eqn:ne1}
  \frac{\mathrm{d}}{\mathrm{d}t}( M_s \bm{U}_c) = \oint_{\partial V_s} (-p \bnhat + \bm{\uptau} \bcdot \bnhat ) \, \mathrm{d}A ,
\end{equation}
\begin{equation}\label{eqn:ne2}
    \frac{\mathrm{d}}{\mathrm{d}t}(\tens{I} \bm{\Omega}_c) = \oint_{\partial V_s}\bm{r} \times (-p \bnhat + \bm{\uptau} \bcdot \bnhat ) \, \mathrm{d}A.
\end{equation}
Here, $M_s(t)$ is the mass of the solid which is retained in the time derivative owing to melting, and $\tens{I}(t)$ is the solid object's inertia tensor, which again, varies with time due to melting. On the right-hand side of \eqref{eqn:ne1} and \eqref{eqn:ne2} are the hydrodynamic loads and torques exerted on the solid respectively. These are expressed as an integration over the solid body interface, denoted $\partial V_s$, and are determined by the surface pressures and viscous stresses on the solid surface where $\bm{\uptau} \equiv \mu (\bnabla \bm{u} + \bnabla \bm{u}^\mathrm{T})$ is the viscous stress tensor, $\mu \equiv \rho \nu$ is the dynamic viscosity, and $\bnhat$ is the normal vector pointing into the fluid domain. Here, we restrict discussion to systems with single solid objects, such that collision forces and torques which could potentially appear on the right-hand side of \eqref{eqn:ne1} and  \eqref{eqn:ne2} are absent. 

At the solid-liquid interface location, which we denote $\bm{X}$, the liquid velocity obeys a no-slip boundary condition.
Here, there exists a kinematic motion of the interface (i.e. translation and rotation) following from the Newton--Euler equations, but also a thermodynamically-driven motion of the interface due to melting, governed by the Stefan condition \citep{davis2001}.
This thermodynamic interface motion, however, does not alter the motion of fluid adjacent to it and consequently, the no-slip condition is applied with respect to only the kinematic interface velocity:
\begin{equation}\label{eq:goveqn_bc_u1}
        \bm{u}(\bm{x} = \bm{X}) \equiv \bm{U} = \bm{U}_c + \bm{\Omega}_c \times \bm{r}
\end{equation}
where $\bm{r} = \bm{X} - \bm{X}_c$.
In the absence of an external flow or solid motion, the position of the solid-liquid interface moves at a velocity $\mathrm{d}\bm{X}/\mathrm{d}t=\bm{U}_\mathrm{melt}$ as the solid melts.
This interface melt velocity is normal to the interface and the physical law which governs melting at a solid-liquid boundary is given by the Stefan condition, which states that the interface heat fluxes drive the interface melt velocity, $\bm{U}_{\mathrm{melt}}$ \citep{carslaw1959,ozicsik1980,alexiades1993,worster2000,davis2001}:
\begin{equation}\label{eq:stefan_condition}
\frac{\mathcal{L}}{c_p} \bm{U}_{\mathrm{melt}} =  \left( \kappa \frac{\partial \theta}{\partial n} \bigg |_{\mathrm{solid}} - \kappa \frac{\partial \theta}{\partial n} \bigg |_{\mathrm{liquid}}\right) \mathbf{\hat{n}}
\end{equation}
where $\mathcal{L}$ is the latent heat of fusion associated with the solid phase, $c_p$ is the specific heat capacity, and $\partial / \partial n \equiv \bnhat \bcdot \nabla$ denotes a gradient in the local normal direction. Note that \eqref{eq:stefan_condition} is a simplified form of the Stefan condition having assumed matched thermal diffusivities, $\kappa$, and specific heats $c_p$ between the solid and liquid phase. Typically, the Stefan condition is characterised by the Stefan number, $St \equiv \mathcal{L} / (c_p \Delta \theta)$ where $\Delta \theta$ is a characteristic temperature difference in the physical system. 

For temperature, the boundary condition to be satisfied at the interface is
\begin{equation}\label{eqn:bc_tmelt}
    \theta(\bm{x} = \bm{X}) = \theta_{\mathrm{melt}}.
\end{equation}
where $\theta_{\mathrm{melt}}$ is the constant melt temperature. This neglects Gibbs--Thomson effects where the local curvature of the interface may affect the local melt temperature, $\theta_{\mathrm{melt}}$ \citep{alexiades1993,davis2001}. In practice, Gibbs--Thomson effects in melting problems typically occur at the nanoscale, such that they may be neglected \citep{buffat1976}. 
\section{Numerical methods}\label{sec:numerical}
\subsection{Eulerian flow solver}
We solve the incompressible Navier--Stokes equations alongside the temperature equations (\ref{eq:goveqn1}--\ref{eq:goveqn_tsolid}) using a staggered, second-order finite difference discretisation. A uniform mesh spacing, $\Delta$, is adopted for all directions in the present work, although the same method has already been used with nonuniform meshes in prior work \citep{vanella2009,van2015,detullio2016}. The incompressible Navier--Stokes equations are solved using a fractional-step procedure \citep{chorin1968,kim1985} to enforce incompressibility. The convective terms for both temperature and velocity are discretised explicitly using the low-storage, 3rd order Runge--Kutta (RK3) method detailed in \citep{rai1991}, while diffusive terms are treated semi-implicitly using a Crank--Nicholson scheme. In discrete form, the equations may be written as:
\begin{align}\label{eq:goveqn_discrete1}
\frac{\bm{\hat{u}} - \bm{u}^n}{ \Delta t} &= - \alpha_n \mathcal{G}(p^n) + \gamma_n \mathcal{H}^n + \zeta_n \mathcal{H}^{n-1} + \nu \frac{\alpha_n}{2} \left[ \mathcal{D}(\bm{\hat{u}}) + \mathcal{D}(\bm{u}^n) \right],  \\
\dfrac{\theta^{*} - \theta^n}{ \Delta t} &= \gamma_n \mathcal{H}^n_\theta + \zeta_n \mathcal{H}^{n-1}_\theta + \kappa \frac{\alpha_n}{2} \left[ \mathcal{D}(\theta^{*}) + \mathcal{D}(\theta^n) \right]. \label{eq:goveqn_discrete2}
\end{align}
Here, the subscripts and superscripts $n$ denote a value at each RK3 substep, $n$, with the coefficients $\gamma_n = [8/15,5/12,3/4]$, $\zeta_n = [0,-17/60,-5/12]$, $\alpha_n = \gamma_n + \zeta_n$, $\Delta t$ is the time step, $\mathcal{H}$, $\mathcal{H}_\theta$, contain the convective terms, $\mathcal{G}(\cdot)$ represents a discrete gradient operator, and $\mathcal{D}(\cdot)$ represents a discrete Laplacian operator. The convective terms, $\mathcal{H}$, $\mathcal{H}_\theta$ are discretised using a standard skew-symmetric form for staggered grids. We also account for the temperature equation in the solid at this same step (equation \ref{eq:goveqn_tsolid}) by appropriately modifying $\mathcal{H}_\theta$ to be a discrete approximation of $\bnabla \bcdot (\bm{u} \theta)$ in fluid cells or $\bnabla \bcdot (\bm{U}_s \theta)$ in solid cells where $\bm{U}_s$ follows from equation \eqref{eq:Usolid}. This entails tagging each computational cell as either fluid, solid, or interface cells at each timestep and is achieved using a ray-tagging procedure described in \citep{orourke1998}. The velocity and temperature $\bm{\hat{u}}$, $\theta^*$, are intermediate quantities to be further updated when advancing to the next RK3 substep. Stable time-integration with the present RK3 scheme can be achieved for Courant--Friedrichs--Lewy numbers in the range $\mathrm{CFL} \equiv \text{max}_i(|u_i| \Delta t / \Delta x_i) < \sqrt{3}$ considering only the flow solver. Direct-forcing IBMs as in the present work do not impose significant timestepping restrictions on top of this \citep{fadlun2000,uhlmann2005}. Melting via the Stefan condition \eqref{eq:stefan_condition}, too, does not tend to impose serious restrictions on the timestep owing to the typically-slow time scales associated with melting relative to the flow timescales. Setting $\delta \bm{\hat{u}} \equiv \bm{\hat{u}} - \bm{u}^n$, $\delta \theta \equiv \theta^{*} - \theta^n$, $\beta \equiv \nu \alpha_n \Delta t/2  $, $\beta^\theta \equiv  \kappa \alpha_n \Delta t /2$, equations (\ref{eq:goveqn_discrete1}--\ref{eq:goveqn_discrete2}) may be re-cast into the form:
\begin{align}\label{eq:goveqn_discrete3}
    (1 - \beta\mathcal{D})\delta \bm{\hat{u}} &= - \alpha_n \Delta t\mathcal{G}(p^n) + \gamma_n \Delta t \mathcal{H}^n + \zeta_n \Delta t\mathcal{H}^{n-1} + 2 \beta  \mathcal{D}(\bm{u}^n), \\
    (1 - \beta^\theta\mathcal{D})\delta \theta &= \gamma_n \Delta t \mathcal{H}^n_\theta + \zeta_n \Delta t\mathcal{H}^{n-1}_\theta + 2 
 \beta^\theta  \mathcal{D}(\theta^n). \label{eq:goveqn_discrete4}
\end{align}
Solutions to equations (\ref{eq:goveqn_discrete3}--\ref{eq:goveqn_discrete4}) require inversion of large sparse matrices. This is circumvented by adopting the so-called approximate factorisation or alternating direction implicit technique \citep{peaceman1955,ferziger2002,moin2010}, which reduces the systems (\ref{eq:goveqn_discrete3}--\ref{eq:goveqn_discrete4}) to a sequence of three tri-diagonal system inversions. Following the solution to (\ref{eq:goveqn_discrete3}--\ref{eq:goveqn_discrete4}), the IB direct forcing is added:
\refstepcounter{equation}
$$
\bm{u}^* = \bm{\hat{u}} + \alpha_n \Delta t \bm{f}, \quad \theta^{n+1} = \theta^* + \alpha_n \Delta t f^\theta
\eqno{(\theequation{a,b})}\label{eq:ib_update}
$$
where precise calculations of $\bm{f}$, $f^\theta$ will be detailed in \S \ref{eq:mls-ibm}. Finally the projection step of the fractional-step is performed to obtain a solenoidal velocity field at the next RK3 substep, $\bnabla \bcdot \bm{u}^{n+1} = 0$, through
\begin{equation}\label{eq:press_proj}
    \bm{u}^{n+1} = \bm{u}^* - \alpha_n \Delta t \bnabla \varphi
\end{equation}
where $\varphi$ is a solution to the Poisson equation $\nabla^2 \varphi = \bnabla \bcdot \bm{u}^* / (\alpha_n \Delta t)$. The pressure at timestep $n+1$ is updated as $p^{n+1} = p^n + \varphi - \beta\nabla^2 \varphi$.
\subsection{Moving least-squares immersed boundary direct-forcing}\label{eq:mls-ibm}
\begin{figure}
\centering
		\includegraphics[]{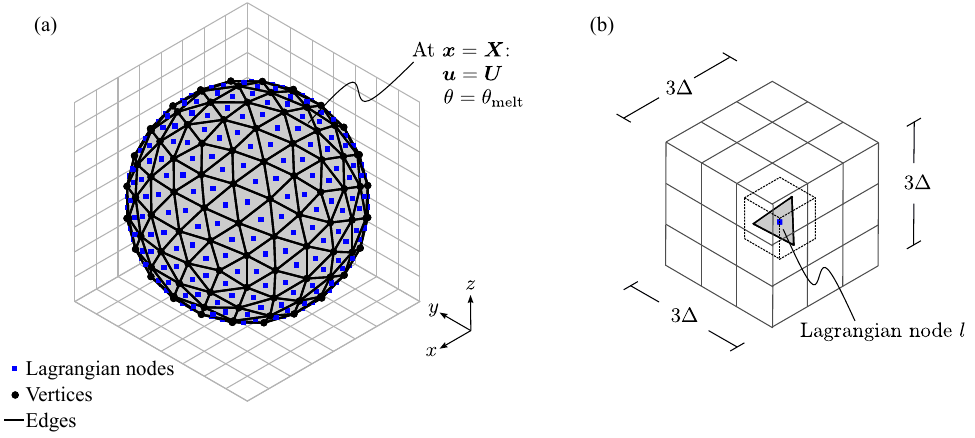}
 	\caption{(a) Discrete formulation of our present method. The solid object is immersed in a fixed Cartesian domain and the solid object is discretised as a triangulated surface comprised of interconnected vertices, edges, and triangular faces. Lagrangian nodes are situated at triangle centroids, and correspond to discrete locations at which the surface boundary conditions $\bm{u}=\bm{U}, \theta = \theta_{\mathrm{melt}}$ are enforced. (b) The local support domain of Eulerian cells for a single Lagrangian node, $l$. This support domain comprises $3 \times 3\times 3$ Eulerian cells surrounding the closest neighbouring Eulerian cell to the Lagrangian node (dashed-line cube). Interpolation within this support domain is performed to evaluate the IB direct-forcing.}
 	\label{fig:fig1_formulation}
\end{figure}
Here, we will recall the procedure of evaluating the IB direct-forcing, $\bm{f}$, $f^\theta$, required in (\ref{eq:ib_update}a,b), which will serve to enforce the boundary conditions \eqref{eq:goveqn_bc_u1} and \eqref{eqn:bc_tmelt} at the interface. This procedure entails interpolation operations between the Eulerian fields and a discrete representation of the Lagrangian solid mesh, and is achieved using the MLS interpolation \citep{lancaster1981,liu2005,vanella2009,detullio2016,viola2020}. In this framework, the immersed solid object is discretised as a triangulated surface as illustrated in figure \ref{fig:fig1_formulation}(a) and is composed of faces, edges, and vertices. Lagrangian nodes are situated at triangle centroids and correspond to discrete locations at which the IB direct-forcing is applied. Denoting the spatial location of Lagrangian node $l$ as $\bm{X}_l \equiv [X_l,Y_l,Z_l]^{\mathrm{T}}$, we interpolate a quantity $q$ from the Eulerian field to $\bm{X}_l$ and denote this $Q_l \equiv q(\bm{X}_l)$. An MLS interpolation evaluates this as:
\begin{equation}\label{eq:mls-equation1}
	 Q_l \approx \mathbf{p}^{\mathrm{T}}(\bm{X}_l) \mathbf{a}(\bm{X}_l) = \sum_{j=1}^{m=4} p_j(\bm{X}_l) a_j(\bm{X}_l).
\end{equation}
Here, $q$ will represent either the three components of velocity, pressure, or temperature in the Eulerian mesh following the solutions to equations (\ref{eq:goveqn_discrete1}--\ref{eq:goveqn_discrete2}), i.e. either of $\{\hat{u},\hat{v},\hat{w},p,\theta^*\}$. The vector $\mathbf{p}^{\mathrm{T}}(\bm{X}_l) = [1,X_l,Y_l,Z_l]$ is a linear shape function, and $\mathbf{a}(\bm{X}_l)$ is a vector of coefficients obtained by solving a local weighted least-squares problem at $\bm{X}_l$. That is, we minimise a weighted L2-norm, $J$ \citep{liu2005}:
\begin{equation}
    J = \sum^{N_e = 27}_{k=1}\mathcal{W}(\bm{X}_l-\bm{x}_k)[\mathbf{p}^{\mathrm{T}}(\bm{x}_k)\mathbf{a}(\bm{X}_l) - q_k ]^2
\end{equation}
where $q_k \equiv q(\bm{x}_k)$ denotes the discrete values of $q$ at Eulerian cell $k$ and $\mathcal{W}$ is a prescribed analytic weight function. The summation is performed over the $N_e = 27$ neighbouring Eulerian cells closest to the Lagrangian node, which corresponds to a local $3 \times 3 \times 3$ interpolation support domain as illustrated in figure \ref{fig:fig1_formulation}(b) and is the typical choice for MLS IB forcing schemes \citep{vanella2009,detullio2016,spandan2017,viola2020}. The coefficients $\mathbf{a}(\bm{X}_l)$ are obtained by minimising $J$ with repsect to $\mathbf{a}$, $\partial J / \partial \mathbf{a} = \mathbf{0}$. This leads to linear systems of the form \citep{liu2005}:
\begin{equation}\label{eq:mls-equation2}
    \tens{A}(\bm{X}_l) \mathbf{a}(\bm{X}_l) = \tens{B}(\bm{X}_l) \mathbf{q}
\end{equation}
where $\mathbf{q} = [q_1,q_2,\dots,q_{N_e}]^{\mathrm{T}}$ is a vector containing the $N_e$ values of $q$ in the support domain. The matrices $\tens{A}$ and $\tens{B}$ are given by:
\begin{equation}
    	\tens{A}(\bm{X}_l) = \sum_{k=1}^{N_e = 27} \mathcal{W}(\bm{X}_l-\bm{x}_k)\mathbf{p}(\bm{x}_k)\mathbf{p}^{\mathrm{T}}(\bm{X}_l),
\end{equation}
\begin{equation}
	\tens{B}(\bm{X_l}) = \begin{bmatrix}
		\mathcal{W}(\bm{X}_l-\bm{x}_1)\mathbf{p}(\bm{x}_1) & \mathcal{W}(\bm{x}-\bm{x}_2)\mathbf{p}(\bm{x}_2) & \dots & \mathcal{W}(\bm{X}_l-\bm{x}_{N_e})\mathbf{p}(\mathbf{x}_{N_e})
\end{bmatrix}.
\end{equation}
For the present $m=4$, $N_e = 27$, the sizes of $\tens{A}$ and $\tens{B}$ are $4 \times 4$ and $4 \times 27$ respectively. From \eqref{eq:mls-equation2} we have $\mathbf{a}(\bm{X}_l) = \tens{A}^{-1}(\bm{X}_l)\tens{B}(\bm{X}_l)\mathbf{q}$. Substituting this result into right-hand side of \eqref{eq:mls-equation1} yields the final expression from which $Q_l$ is computed:
\begin{equation}\label{eq:mls-interpeqn}
    Q_l \approx \mathbf{p}^{\mathrm{T}}\tens{A}^{-1}(\bm{X}_l)\tens{B}(\bm{X}_l)\mathbf{q} \equiv \mathbf{\Phi}^\mathrm{T} \mathbf{q} = \sum_{k=1}^{N_e = 27} \phi^l_k(\bm{X}_l) q_k
\end{equation}
where we have defined $\mathbf{\Phi}^\mathrm{T} \equiv \mathbf{p}^{\mathrm{T}}(\bm{X}_l)\tens{A}^{-1}(\bm{X}_l)\tens{B}(\bm{X}_l)$. Thus, for each Lagrangian node $l$, the elements of $\mathbf{\Phi}$, $\phi^l_k$, represent the weighted contribution of the Eulerian cell value $q_k$ to $Q_l$. For the weight function, $\mathcal{W}$, we use the Gaussian distribution, a common choice for MLS interpolation \citep{liu2005}:
\begin{equation}
    \mathcal{W}(\bm{X}_l - \bm{x}_k) = 
    \left\{ \begin{aligned}
    & \mathrm{e}^{-(d_k/\varepsilon)^2}, && d_k \leq 1, \\
    &0, && d_k > 1
    \end{aligned} \right.
\end{equation}
where $\varepsilon = 0.3$, $d_k \equiv |\bm{X}_l - \bm{x}_k| / d_i$ and $d_i = 3 \Delta/2$ is the half-width of the support domain in each direction, $i=\{1,2,3\}$, which remains fixed for the present uniform grid (figure \ref{fig:fig1_formulation}b).

With $Q_l$ interpolated, we now evaluate the direct-forcing associated with each Lagrangian node $l$ required to impose the boundary value at the Lagrangian node, $Q^b_l$. Each Lagrangian node $l$ has a discrete forcing of the form:
\begin{equation}\label{eq:fibm_q}
    F_l^Q = \frac{Q^b_l - Q_l}{\alpha_n \Delta t}.
\end{equation}
For velocities, $Q^b_l$ is comprised of the translational and rotational velocities as in \eqref{eq:goveqn_bc_u1}, whereas for temperature, $Q^b_l$ is simply the fixed melting temperature $\theta_{\mathrm{melt}}$ as in (\ref{eqn:bc_tmelt}). Thus, for velocity and temperature respectively, \eqref{eq:fibm_q} is:
\begin{equation}
    \bm{F}_l = \frac{\bm{U}_l^b - \bm{\hat{U}}_l}{\alpha_n \Delta t}, \quad F^\theta_l = \frac{\theta_{\mathrm{melt}}- \theta^*_l}{\alpha_n \Delta t}
\end{equation}
where in accordance with the boundary condition \eqref{eq:goveqn_bc_u1}, the velocity of Lagrangian node $l$ to be enforced, $\bm{U}^b_l$ is:
\begin{equation}\label{eqn:ibm-ubl}
    \bm{U}^b_l = \bm{U}_c + \bm{\Omega}_c \times \bm{r}_l
\end{equation}
with $\bm{r}_l \equiv \bm{X}_l - \bm{X}_c$. To satisfy the boundary conditions in the Eulerian fields with (\ref{eq:ib_update}a,b), the discrete Lagrangian forcing, $\bm{F}_l$, $F_l^\theta$ must be transferred to their Eulerian counterparts, $\bm{f}$, $f^\theta$, through a spreading operation. For each Eulerian cell $k$, within the support domain of Lagrangian node $l$ (figure \ref{fig:fig1_formulation}b), its associated Eulerian direct-forcing is evaluated as:
\begin{equation}\label{eq:fibm_discrete}
    \bm{f}_k = \sum_{l\in \mathcal{N}_k}c_l \phi^l_k \bm{F}_l, \quad f^\theta_k = \sum_{l\in \mathcal{N}_k}c_l \phi^l_k F_l^\theta.
\end{equation}
The summation occurs over all $l$ Lagrangian nodes for which Eulerian cell $k$ is a cell within the support domain of the Lagrangian node, denoted by $l \in \mathcal{N}_k$. The coefficient $c_l$ is a transfer coefficient associated with each Lagrangian node. As demonstrated by \citet{vanella2009}, the choice $c_l = V_l / V_E$ guarantees discrete conservation of total torque and momentum for uniform Eulerian grid spacings when passing from the Lagrangian mesh to Eulerian mesh, where $V_E = \Delta^3$ is the Eulerian mesh volume, $V_l = A_l \Delta$ is the volume associated with Lagrangian marker $l$ and $A_l$ is the area of triangle $l$. For a uniform mesh, $c_l = A_l / \Delta^2$, which for rigid bodies would remain constant. When melting however, the solid-body shrinks with time such that the triangle areas $A_l$ will vary with time. Consequently, $c_l$ should be updated at each timestep.

Finally, with $\bm{f}$, $f^\theta$ obtained from \eqref{eq:fibm_discrete}, the Eulerian velocity and temperature fields may be updated according to (\ref{eq:ib_update}a,b), $\bm{u}^* = \bm{\hat{u}} + \alpha_n \Delta t \bm{f}, \theta^{n+1} = \theta^* + \alpha_n \Delta t f^\theta$, to satisfy the boundary conditions at the solid-liquid interface. Note for velocity however, that a further update in the form of the pressure correction step (\ref{eq:press_proj}) is applied to $\bm{u}^*$ to obtain $\bm{u}^{n+1}$ since $\bm{u}^*$ will not be divergence-free in general. Consequently, this means that $\bm{u}^{n+1}$ does not exactly satisfy the interface boundary conditions, and there will instead be residual velocities near the interface. Simultaneously enforcing incompressibility and the surface boundary conditions is one of the primary difficulties of direct-forcing IBMs \citep{verzicco2023}. Schemes have been proposed to either minimise \citep{fadlun2000} or eliminate these residuals \citep{kim2001,taira2007} with varying degrees of complexity and increase in computational cost. These are not considered in the present work.

Concerning parallelisation strategies, the treatment of mixed Eulerian--Lagrangian mesh problems such as the present work can pose a problem as the underlying parallelisation schemes for solutions on Eulerian and Lagrangian meshes can greatly differ. Following prior work using the MLS-IBM, our present work adopts the parallelisation scheme described in \S 5 of \citet{spandan2017} and is briefly summarised as follows. A distributed memory paradigm is used for the flow solver following the Message Passing Interface (MPI) standard, where the Cartesian Eulerian domain is decomposed into 1D `slabs' and each processor is designated an individual slab for performing Eulerian computations associated with the spatial coordinates of the slab. Such a decomposition strategy and its associated implementation is well-known for incompressible flow solvers, although alterative decomposition strategies exist \citep{van2015}. The parallelisation of the Lagrangian computations can pose a challenge, especially in the present case where the solid object is free to cross between spatial regions contained by different MPI processes. Following \citet{spandan2017}, this difficulty is circumvented by storing all the triangulated geometry information on every processor concurrently. Crucially however, the costliest operations in the MLS-interpolations for each Lagrangian node are only performed by the unique processor containing the spatial coordinates of the Lagrangian node. Interpolation from the $3 \times 3 \times 3$ support domain is naturally handled in a second-order finite difference framework, as a layer of halo/ghost cells is used on both sides of a slab processor, guaranteeing that the processor will have the sufficient 3-layer thick Eulerian cell information of the support domain. We note that this same parallelism is employed for the MLS-interpolation when probe locations (\S \ref{sec:fsi}--\ref{sec:stefan}, figure \ref{fig:fig2_mlsProbe}b) are considered as opposed to the Lagrangian node locations.
\subsection{Fluid-solid interaction solver}\label{sec:fsi}
\begin{figure}
\centering
		\includegraphics[]{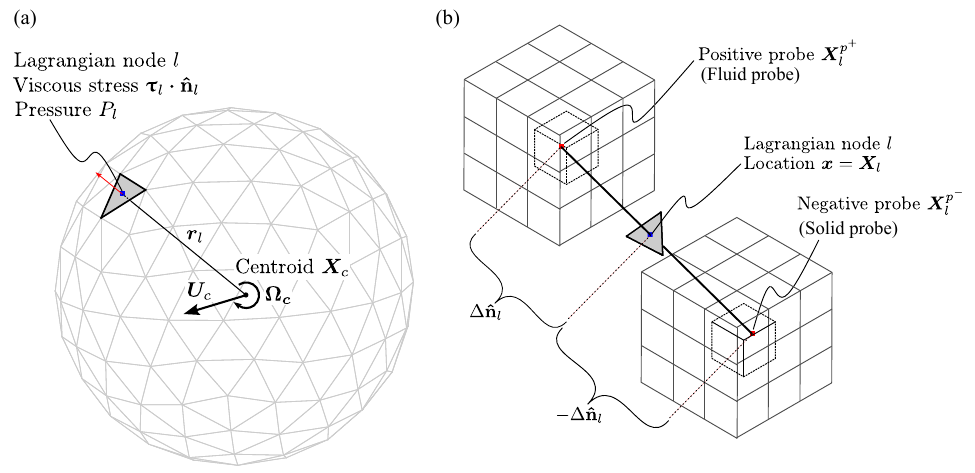}
 	\caption{(a) Illustration of the FSI-solver process. The Newton--Euler equations are solved for the angular and translational centroid velocities, $\bm{\Omega}_c$, $\bm{U}_c$ by accumulating contributions from each Lagrangian node. (b) Sketch of the normal probes for each Lagrangian node $l$, used to evaluate the hydrodynamic loads (\S \ref{sec:fsi}) and surface heat-fluxes (\S \ref{sec:stefan}). The lengths of the probes have been exaggerated for visual clarity.}
 	\label{fig:fig2_mlsProbe}
\end{figure}

Rotation and translation of the solid object is accomplished through time-integration of the Newton--Euler equations (\ref{eqn:ne1}--\ref{eqn:ne2}), which requires computation of surface-integrated hydrodynamic loads, $\bm{L}_H$, and torques, $\bm{T}_H$, of the form $\oint_{\partial V_s}(\cdot)\mathrm{d}A$. We discretise these surface integrals based on a summation over all $N_l$ triangles of the solid object (figure \ref{fig:fig2_mlsProbe}a):
\begin{align}\label{eq:ne-mls1}
\bm{L}_H \equiv \oint_{\partial V_s} (-p \bnhat + \bm{\uptau} \bcdot \bnhat ) \, \mathrm{d}A &\approx \sum_{l=1}^{N_l}(-P_l \bnhat_l + \bm{\uptau}_l \bcdot \bnhat_l)A_l, \\
\bm{T}_H \equiv \oint_{\partial V_s} \bm{r} \times (-p \bnhat + \bm{\uptau} \bcdot \bnhat ) \, \mathrm{d}A &\approx \sum_{l=1}^{N_l}\bm{r}_l \times (-P_l \bnhat_l + \bm{\uptau}_l \bcdot \bnhat_l)A_l \label{eq:ne-mls2}
\end{align}
where $P_l$, $\bm{\uptau}_l$ is the pressure and viscous stress tensor at Lagrangian node $l$ respectively. Following prior studies which have adopted MLS-interpolation for FSI problems, $P_l$ and $\bm{\uptau}_l$ are obtained through an extrapolation from probe locations \citep{vanella2009,viola2020}. A schematic illustration is presented in figure \ref{fig:fig2_mlsProbe}(b). Here, outward (positive) and inward (negative) probe locations for each Lagrangian node are obtained by extending the Lagrangian node location in the local normal direction with a spacing equal to the Eulerian mesh spacing, $\pm \Delta \bnhat_l$. That is, $\bm{X}^{p^{\pm}}_l \equiv \bm{X}_l \pm \Delta \bnhat_l$ where $\bm{X}^{p^{\pm}}_l$ denotes the location of a positive or negative probe. Since evaluation of $\bm{L}_H$, $\bm{T}_H$ in (\ref{eq:ne-mls1}--\ref{eq:ne-mls2}) only requires outward-pointing normal information, only information at positive probes $\bm{X}^{p^+}_l$ will be used for evaluating $\bm{L}_H$, $\bm{T}_H$. The negative, inward-pointing probes will be used later when we consider the implementation of melting (\S \ref{sec:stefan}). The pressures $P_l$ and velocity gradients $\partial \bm{u} / \partial x_i$ from which $\bm{\uptau}_l$ are built are obtained from values at $\bm{X}_l^{p^+}$ \citep{vanella2009,viola2020}:
\begin{equation}\label{eq:mls-probe1}
P_l \equiv p(\bm{X}_l) \approx p(\bm{X}_l^{p^+}) - \Delta \frac{\partial p}{\partial n} \bigg |_{\bm{X}_l^{p^+}},
\end{equation}
\begin{equation}\label{eq:mls-probe2}
    \frac{\partial \bm{u}}{\partial x_i} \bigg |_{\bm{X}_l} \approx \frac{\partial \bm{u}}{\partial x_i} \bigg |_{\bm{X}^{p^+}_l}.
\end{equation}
Here, \eqref{eq:mls-probe1} is obtained from an $O(\Delta^2)$ Taylor series expansion, while \eqref{eq:mls-probe2} corresponds to a boundary-layer approximation near a solid interface. Values at $\bm{X}_l^{p^+}$ are obtained through an MLS interpolation as with equation \eqref{eq:mls-interpeqn}:
\begin{equation}
    p(\bm{X}^{p^+}_l) = \sum_{k=1}^{N_e = 27} \phi^l_k(\bm{X}^{p^+}_l) p_k,
\end{equation}
\refstepcounter{equation}
$$
\frac{\partial \bm{u}}{\partial x_i} \bigg |_{\bm{X}_l^{p^+}} = \sum_{k=1}^{N_e = 27} \frac{\partial \phi^l_k}{\partial x_i}\bigg|_{\bm{X}_l^{p^+}}\bm{u}_k, \qquad \frac{\partial p}{\partial x_i} \bigg |_{\bm{X}_l^{p^+}} = \sum_{k=1}^{N_e = 27} \frac{\partial \phi^l_k}{\partial x_i}\bigg|_{\bm{X}_l^{p^+}}p_k.
\eqno{(\theequation{a,b})}\label{eq:mls-gradients}
$$
As described by \citet{liu2005}, one can determine the derivatives of the MLS-shape functions $\partial \phi^l_k / \partial x_i$ analytically. The normal-projection of the pressure gradient in expression \eqref{eq:mls-probe1} is simply the dot product of $\bnhat_l$ with (\ref{eq:mls-gradients}b), $\partial p / \partial n \equiv  \bnhat \cdot \nabla p$. Prior studies adopting the MLS interpolation for FSI problems have typically approximated the normal pressure gradient through a boundary-layer approximation $\partial p / \partial n|_{\bm{X}_l} \approx - (\mathrm{D}\bm{U}_l/\mathrm{D}t) \bcdot \bnhat_l$ where $\mathrm{D}\bm{U}_l/\mathrm{D}t$ is the acceleration of Lagrangian node $l$ \citep{yang2006,vanella2009}. We do not adopt this approach here, instead evaluating the pressure gradients with an MLS-interpolation as with (\ref{eq:mls-gradients}b).

With a scheme for evaluating $\bm{L}_H$, $\bm{T}_H$ established, we integrate the Newton--Euler equations in time to obtain the centroid velocity and angular velocity at the next timestep, $\bm{U}_c^{n+1}$, $\bm{\Omega}_c^{n+1}$ respectively:
\refstepcounter{equation}
$$
(M_s \bm{U}_c)^{n+1} = (M_s \bm{U}_c)^{n} + \gamma_n \Delta t \bm{L}_H^n + \zeta_n \Delta t \bm{L}_H^{n-1}, \quad   (\tens{I}\bm{\Omega_c})^{n+1}  = (\tens{I}\bm{\Omega_c})^{n} + \gamma_n \Delta t \bm{T}_H^n + \zeta_n \Delta t \bm{T}_H^{n-1}. 
\eqno{(\theequation{a,b})}\label{eq:mls-ne1}
$$
Here, we retain the same solid mass between timesteps $n$ and $n+1$: $M_s^{n+1} \approx M_s^{n}$ in (\ref{eq:mls-ne1}a), which effectively treats the solid mass as invariant during the solid-body translation step. Such an assumption is applicable provided the rate-of-change of linear momentum due to mass-loss, $\mathrm{d}M_s/\mathrm{d}t$ is a negligible contribution to the total rate-of-change of linear momentum: $\bm{U}_c (\mathrm{d}M_s/\mathrm{d}t) \ll M_s (\mathrm{d}\bm{U}_c/\mathrm{d}t)$ such that $\mathrm{d}(M_s \bm{U}_c) / \mathrm{d}t \approx M_s (\mathrm{d}\bm{U}_c/\mathrm{d}t)$.
This assumption is adequate for most practical melting problems in which the solid phase has a relatively large latent heat of fusion such as ice-water systems. A higher latent of fusion entails a larger Stefan number and 
a large Stefan number (coupled to a large Pecl\'{e}t number) implies that the timescale of melting, which sets the magnitude of $\mathrm{d}M_s/\mathrm{d}t$, is much slower than the timescales of motion in the system.
For (\ref{eq:mls-ne1}b), the inertia tensor $\tens{I}(t)$ must be updated at every timestep $n$ since the solid geometry dynamically changes owing to melting. The inertia tensor is a $3 \times 3$ symmetric tensor and is defined as:
\begin{equation}
    \tens{I}(t) \equiv 
    \rho \begin{bmatrix}
         \int_{V_s}(y^{\prime 2} + z^{\prime 2}) \, \mathrm{d}V & -\int_{V_s}x^{\prime} y^{\prime} \, \mathrm{d}V & -\int_{V_s}x^{\prime} z^{\prime} \, \mathrm{d}V \\[6pt]
         \dots & \int_{V_s}(x^{\prime 2} + z^{\prime 2}) \, \mathrm{d}V & -\int_{V_s}y^{\prime} z^{\prime} \, \mathrm{d}V \\[6pt]
         \dots & \dots  & \int_{V_s}(x^{\prime 2} + y^{\prime 2}) \, \mathrm{d}V
    \end{bmatrix}
\end{equation}
where $(x^\prime,y^\prime,z^\prime)$ are solid-body coordinates relative to the solid's centroid. We compute the inertia tensor at timestep $n$, $\tens{I}^n$, using an efficient numerical integration scheme described by \citet{kallay2006} for triangulated surfaces, where the volume integrals defining $\tens{I}$ are computed as a sum of signed tetrahedral volume integrals \citep{lien1984}. This same scheme is also used to compute the solid mass at each timestep, $M_s(t) \equiv \rho \int_{V_s}\, \mathrm{d}V$.

As noted by \citet{ardekani2016} an iterative approach to solve (\ref{eq:mls-ne1}b) for the angular velocity at the new timestep, $\bm{\Omega}_c^{n+1}$ is required since the inertia tensor $\tens{I}^{n+1}$, too, is an unknown to be determined. We follow the procedure of \citet{ardekani2016} which is as follows:
\begin{enumerate}
    \item Set an initial guess for $\tens{I}^{n+1}$ as $\tens{I}^n$, call this $\tens{I}^* = \tens{I}^n$.
    \item Compute $\bm{\Omega}_c^{n+1}$ using (\ref{eq:mls-ne1}b) with the provisional inertia-tensor estimate $\tens{I}^* = \tens{I}^{n+1}$.
    \item The rotation of (\ref{eq:mls-ne1}b) coincides with a rotation about an axis $\mathbf{s} = (\bm{\Omega}_c^{n} + \bm{\Omega}_c^{n+1})/2$ and angle $\Theta = |(\bm{\Omega}_c^{n} + \bm{\Omega}_c^{n+1})/2|(\alpha_n \Delta t)$. A corresponding rotation matrix representation, $\tens{R} = \tens{R}(\mathbf{s},\Theta)$ can be computed \citep{diebel2006}. This is used to update the inertia tensor estimate $\tens{I}^*_{\mathrm{new}} = \tens{R}\tens{I}^*\tens{R}^{\mathrm{T}}$.
    \item Repeat step 2 with $\tens{I}^*_{\mathrm{new}}$ until convergence within a desired tolerance is reached, $|\tens{I}^*_{\mathrm{new}} - \tens{I}^*_{\mathrm{old}}| < \epsilon$.
\end{enumerate}
Afterwards, the solid's centroid position is updated:
\begin{equation}
    \bm{X}^{n+1}_c = \bm{X}_c^n + \frac{\alpha_n}{2 \Delta t}(\bm{U}^{n+1}_c + \bm{U}^n_c),
\end{equation}
Followed by the rotational update for each Lagrangian node $l$:
\begin{equation}
    \bm{r}^{n+1}_l = \tens{R}^{\mathrm{T}}\bm{r}^n_l
\end{equation}
where $\bm{r}_l$ is the position of Lagrangian node $l$ relative to the solid's centroid. Here $\tens{R}$ will correspond to the final rotation matrix in the iterative procedure described above. Following this, the velocity of each Lagrangian node is then updated according to \eqref{eqn:ibm-ubl}, $\bm{U}_l^{n+1} = \bm{U}_c^{n+1} + \bm{\Omega}_c^{n+1} \times \bm{r}_l^{n+1}$.
\subsection{Imposing the Stefan condition}\label{sec:stefan}
\begin{figure}
\centering
		\includegraphics[]{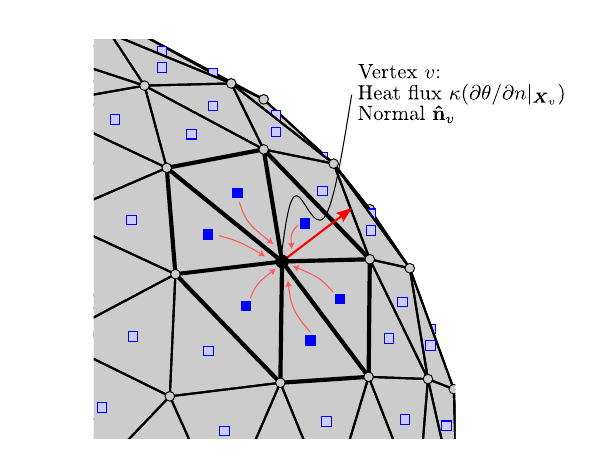}
 	\caption{Interpolation strategy from Lagrangian nodes (triangle centroids) to vertices. Heat-fluxes and normal vectors are interpolated to vertices using information from the neighbouring Lagrangian nodes of each vertex.}
 	\label{fig:fig3_vertNode}
\end{figure}

In this section, we detail the steps for imposing the Stefan boundary condition \eqref{eq:stefan_condition} which governs melting of the interface. For each Lagrangian node $l$, \eqref{eq:stefan_condition} is discretised as:
\begin{equation}\label{eq:stefan-lnode}
    \frac{\mathcal{L}}{c_p}\bm{U}_{\mathrm{melt},l} = \left( \kappa \frac{\partial \theta}{\partial n}\bigg |_{\bm{X}_l,\mathrm{solid}} - \kappa \frac{\partial \theta}{\partial n}\bigg |_{\bm{X}_l,\mathrm{liquid}} \right) \bnhat_l.
\end{equation}
Here, $\bm{U}_{\mathrm{melt},l}$ denotes the melt velocity of Lagrangian node $l$ and $\kappa (\partial \theta / \partial n|_{\bm{X}_l})$ is the one-sided heat-flux at the Lagrangian node. This heat-flux is discontinuous across the solid-facing and liquid-facing side, and is the physical condition which drives melting. Discrete evaluation of these heat-fluxes are achieved by adopting the same MLS-probe framework detailed as with the FSI-solver (\S \ref{sec:fsi}). With reference to figure \ref{fig:fig2_mlsProbe}(b), a positive and negative probe for each Lagrangian node will be associated with the fluid or solid domain. We may approximate the one-sided temperature gradients at $\bm{X}_l$ adopting the same boundary-layer approximation as was done for the viscous stresses in \eqref{eq:mls-probe2}, as well as the MLS-interpolation (\ref{eq:mls-gradients}a,b):
\refstepcounter{equation}
$$
\frac{\partial \theta}{\partial x_i} \bigg |_{\bm{X}_l,\mathrm{solid}} \approx \frac{\partial \theta}{\partial x_i} \bigg |_{\bm{X}_l^{p^-}}, \qquad \frac{\partial \theta}{\partial x_i} \bigg |_{\bm{X}_l,\mathrm{liquid}} \approx \frac{\partial \theta}{\partial x_i} \bigg |_{\bm{X}_l^{p^+}}
\eqno{(\theequation{a,b})}\label{eq:mls-heatfluxes}
$$
\begin{equation}
\frac{\partial \theta}{\partial x_i} \bigg |_{\bm{X}_l^{p^\pm}} = \sum_{k=1}^{N_e = 27} \frac{\partial \phi^l_k}{\partial x_i}\bigg|_{\bm{X}_l^{p^\pm}}\theta_k.
\end{equation}
This would now provide enough information to evaluate the melt velocity for each Lagrangian node, $\bm{U}_{\mathrm{melt},l}$ in \eqref{eq:stefan-lnode}, from which one would update the Lagrangian node locations through $ \mathrm{d}\bm{X}_l / \mathrm{d}t = \bm{U}_{\mathrm{melt},l}$. We do not adopt this approach however. Recall that Lagrangian nodes are situated at triangle centroids. If centroids are to be moved during each timestep, one would be required to recompute the triangulated mesh connectivity at each timestep, which is a non-trivial task. Instead, we presently elect to move the \textit{vertex} coordinates of the Lagrangian mesh, such that the mesh connectivity remains unmodified during this step. Denoting the spatial location of vertex $v$ as $\bm{X}_v$, its position is updated according to:
\begin{equation}\label{eq:stefan-posupdate}
    \bm{X}_v^{n+1} = \bm{X}_v^n + \gamma_n \Delta t \bm{U}_{\mathrm{melt},v}^n + \zeta_n \Delta t \bm{U}_{\mathrm{melt},v}^{n-1}.
\end{equation}
The melt-velocity of vertex $v$, $\bm{U}_{\mathrm{melt},v}$, follows from a vertex-based discretisation of the Stefan condition analogous to \eqref{eq:stefan-lnode}
\begin{equation}\label{eq:stefan-vertex}
        \frac{\mathcal{L}}{c_p}\bm{U}_{\mathrm{melt},v} = \left( \kappa \frac{\partial \theta}{\partial n}\bigg |_{\bm{X}_v,\mathrm{solid}} - \kappa \frac{\partial \theta}{\partial n}\bigg |_{\bm{X}_v,\mathrm{liquid}} \right) \bnhat_v.
\end{equation}
Determination of $\bm{U}_{\mathrm{melt},v}$ requires heat-fluxes and normal vectors at each vertex. These are obtained by interpolating the heat-fluxes and normal vectors of each neighbouring Lagrangian node, the so-called 1-ring neighbourhood of vertex $v$ \citep{botsch2010}. The procedure is illustrated in figure \ref{fig:fig3_vertNode}. Interpolation from geometric faces to vertices is an operation central to computational geometry \citep{jin2005,botsch2010,baerentzen2012}, with various methods available \citep{jin2005}. A robust choice which worked well from our experience is a weighted-interpolation based on the triangle areas, $A_l$:
\refstepcounter{equation}
$$
\mathbf{n}_v = \frac{1}{A_v}\sum_{l \in \mathcal{N}_v} A_l \bnhat_l, \qquad \frac{\partial \theta}{\partial n} \bigg |_{\bm{X}_v} = \frac{1}{A_v}\sum_{l \in \mathcal{N}_v} \frac{1}{3}A_l \frac{\partial \theta}{\partial n}\bigg |_{\bm{X}_l}
\eqno{(\theequation{a,b})}\label{eq:v-to-l-interp}
$$
where $l \in \mathcal{N}_v$ denotes the set of $l$ Lagrangian nodes in the 1-ring neighbourhood of vertex $v$ (figure \ref{fig:fig3_vertNode}) and $A_v = (1/3)\sum_{l \in \mathcal{N}_v} A_l$ is an area associated with vertex $v$, with the $1/3$ weightings corresponding to an equal weighting assignment for each vertex of a given triangle $l$. The vertex normal vector is normalised following interpolation: $\bnhat_v \equiv \mathbf{n}_v / |\mathbf{n}_v|$. Area-weighted face-to-vertex interpolations tend to be a robust choice for accurate normal vector estimations \citep{jin2005}. For heat-flux interpolations, the area-weighted interpolation is a physically-appealing choice, as this choice conserves the surface-integrated heat-flux (i.e. the surface heat transfer) when passing from Lagrangian node indexing, $l$, to vertices, $v$: 
\begin{equation}
\sum^{N_l}_{l=1}A_l \frac{\partial \theta}{ \partial n} \bigg |_{\bm{X}_l} = \sum_{v=1}^{N_v} A_v \frac{\partial \theta}{\partial n} \bigg |_{\bm{X}_v}.
\end{equation}

A clarification is in order regarding our choice to transfer the MLS-interpolated heat-fluxes from Lagrangian nodes to vertices, as opposed to simply performing the MLS-probe interpolations at discrete vertices. This choice is justified on the basis that polyhedral surfaces will in general, have more faces (Lagrangian nodes) than vertices \citep{orourke1998,botsch2010} and consequently, one can utilise more information from the Eulerian mesh at each timestep if a Lagrangian node indexing system, $l$, is used initially. This result can be deduced from Euler's polyhedron identity: $N_v - N_E + N_l = \chi$ where $N_v$, $N_E$, $N_l$ are the number of vertices, edges, and faces of the polyhedron and $\chi$ is the Euler characteristic number of the polyhedron \citep{orourke1998,botsch2010}. Typically, $\chi = O(1)$ for simple polyhedral surfaces and for closed triangulated surfaces, $\chi = 2$, $3N_l = 2N_E$, which results in $N_l \approx 2N_v$ when substituted into Euler's identity for large triangulations \citep{orourke1998}. This availability of more information for Eulerian to Lagrangian mesh interpolation will be particularly advantageous when we consider the dynamic remeshing procedure of our Lagrangian mesh (\S \ref{sec:remesh}). As shall be seen, our remeshing approach includes a procedure in which the Lagrangian mesh is gradually coarsened as the object melts. To reliably melt the solid down to scales approaching the Eulerian grid spacing, ample amounts of feedback-forcing should be maintained between the Eulerian and Lagrangian mesh, which, for a fixed Eulerian grid is most restrictive when the Lagrangian mesh is coarsest. Thus, using the larger number of Lagrangian nodes over vertices helps to meet this condition. 
\section{Dynamic remeshing}\label{sec:remesh}
One of the requirements of front-tracking IB methods is the need for a commensurate Lagrangian and Eulerian mesh spacing to stabilise interpolation between the two grids \citep{prosperetti2009,tryggvason2011}. For a fixed Eulerian grid, such a condition is violated for melting problems where the Lagrangian grid inevitably shrinks with time if no remeshing intervention is applied. Here, we will present our dynamic remeshing procedure for addressing this issue. Briefly, our remeshing algorithm is an incremental procedure comprised of two steps. (1) A coarsening operation based on edge-collapses \citep{garland1997}, which serves to maintain roughly equivalent Eulerian and Lagrangian mesh spacings, followed by (2): a volume-conserving smoothing operation \citep{kuprat2001}, which helps avoid numerically-undesirable features such as skewed triangles or sharp geometric corners from developing, whilst simultaneously enforcing a net-zero volume change in the remeshing step to machine precision. This coupling procedure between coarsening and smoothing is similar to the one described by \citet{roghair2016}, who applied front-tracking for simulating mass transfer in bubbly flows.

Our remeshing procedure is a highly-scalable approach, suitable for simulating high-Reynolds-number turbulence for instance, as it is built upon incremental, local mesh modifications which typically amount to a negligible computational burden per timestep. The trade-off for this advantage will be a gradual loss of solution accuracy as the solid objects melts to smaller sizes. As our remeshing procedure is built on a gradual coarsening procedure, this means that the geometric fidelity, and hence, a loss of solution accuracy is inevitable. The key to dealing with this will be starting with a suitable well-resolved Lagrangian mesh at the initial simulation time, such that at later times, the solution accuracy is not drastically compromised.

\subsection{Mesh coarsening: edge-collapses}\label{sec:ecol}
Here, we will describe our method for coarsening the triangulated geometry. The primary goal of this procedure is to maintain consistency between the fixed Eulerian grid spacing and the Lagrangian grid spacing, which, for a melting (shrinking) object necessitates a coarsening procedure. Our primary operation for this coarsening procedure will be incremental edge-collapse operations and is illustrated in figure \ref{fig:fig4_remesh}(a). Every timestep, we track the edge lengths of our triangulated geometry. Whenever an edge length shrinks below a threshold, the edge is collapsed. We set this threshold to be a fraction of the Eulerian grid spacing, $\leq 0.7 \Delta$, which is a choice based upon prior MLS-IBM literature which found that a Lagrangian grid spacing of approximately $0.7\Delta$ tended to yield the best results \citep{detullio2016,viola2020}. Suppose an edge flagged for collapse is connected by vertices $(\bm{X}_1,\bm{X}_2)$. An edge-collapse converges these vertices to a new, single vertex location $\bm{\overline{X}}$ which we are free to choose. Each edge collapse operation removes 3 edges, 1 vertex, and 2 faces and is a topology-preserving operation. Namely, edge collapses conform to Euler's polyhedral identity $N_v - N_E + N_l = \chi = \mathrm{constant}$ \citep{botsch2010}. This ensures topological artifacts such as surfaces holes are avoided when the mesh connectivity is progressively modified \citep{botsch2010,baerentzen2012}. For determining the new vertex location $\bm{\overline{X}}$, we adopt a popular method in the computational geometry and meshing literature known as Quadric Error Metrics (QEM) \citep{garland1997}. The idea behind QEM is to determine $\bm{\overline{X}}$ as the location which minimises the sum of squared distances to all triangle planes originally neighbouring both $\bm{X}_1$, $\bm{X}_2$. Each triangle $l$ is defined by its plane equation $\bnhat_l \bcdot \bm{x} + d_l = 0$ where $d_l$ is the plane's offset from the coordinate origin and the squared distance to this plane for an arbitrary point $\bm{x}$ is simply $(\bnhat_l \bcdot \bm{x} + d_l)^2$. Thus, QEM computes $\bm{\overline{X}}$ as the location which minimises \citep{garland1998,botsch2010,akenine2018}:
\begin{equation}
\epsilon(\bm{\overline{X}}) = \sum_{ l \in \{ \mathcal{N}_1, \mathcal{N}_2 \} }(\bnhat_l \bcdot \bm{\overline{X}} + d_l)^2. 
\end{equation}
Where we use $l \in \{ \mathcal{N}_1, \mathcal{N}_2 \}$ to denote the set of $l$ triangles neighbouring vertices $\bm{X}_1$ and $\bm{X}_2$. Because $\epsilon(\bm{\overline{X}})$ is quadratic with respect to $\bm{\overline{X}}$ (a \textit{quadric} function in 3D), this can be analytically  minimised, $\partial \epsilon / \partial \bm{\overline{X}} = \mathbf{0}$. This results in a $4 \times 4$ linear system to invert for $\bm{\overline{X}}$ of the form \citep{garland1997}:
\begin{equation}\label{eq:qem_linsystem}
    \begin{bmatrix}
        q_{11} & q_{12} & q_{13} & q_{14} \\
        q_{12} & q_{22} & q_{23} & q_{24} \\
        q_{13} & q_{23} & q_{33} & q_{34} \\
        0 & 0 & 0 & 1
    \end{bmatrix}
    \begin{bmatrix}
        \mbox{%
  \vline height 2ex} \\
        \bm{\overline{\bm{X}}} \\
                \mbox{%
  \vline height 2ex} \\
        1
    \end{bmatrix} = 
    \begin{bmatrix}
        0 \\
        0  \\
        0 \\
        1
    \end{bmatrix}
\end{equation}
Where the elements $q_{ij}$ are built using the triangle-plane equations neighbouring $\bm{X}_1$ and $\bm{X}_2$. Alternative or more compact forms of \eqref{eq:qem_linsystem} are also possible \citep{garland1998,baerentzen2012,akenine2018}.

QEM is favoured as its computation is relatively cheap: we require the solution to a $4 \times 4$ linear system for each edge-collapse using information that is readily available from the geometry (namely, the triangle plane equations from the normal vectors, $\bnhat_l$ and vertex coordinates). The further appeal of this approach is its incremental nature: edge collapses are done locally only where necessary based on the threshold edge length criteria. This is based on a philosophy seeking to keep `artificial' changes to the solid geometry due to interventional remeshing minimal. An incremental approach is also favourable from a computational standpoint compared to methods which might recompute the entire connectivity and coordinates of the Lagrangian mesh at a single timestep. Although the latter may produce higher-quality meshes globally, this is both an expensive and complex computation, especially in 3D. Our incremental edge-collapse approach however, does not pose a significant computational burden per timestep. For example, some of our test cases with $O(10^5)$ triangles only encountered $\lesssim 20$ edge-collapses per timestep at most, entailing a trivial computational load. As an added advantage, the number of Lagrangian nodes will gradually decrease with time as the object melts owing to coarsening, meaning the computational load for Lagrangian--Eulerian operations will only continue to decrease as the object melts.
\subsection{Mesh smoothing: volume-conserving edge-relaxations}
\begin{figure}
\centering
		\includegraphics[]{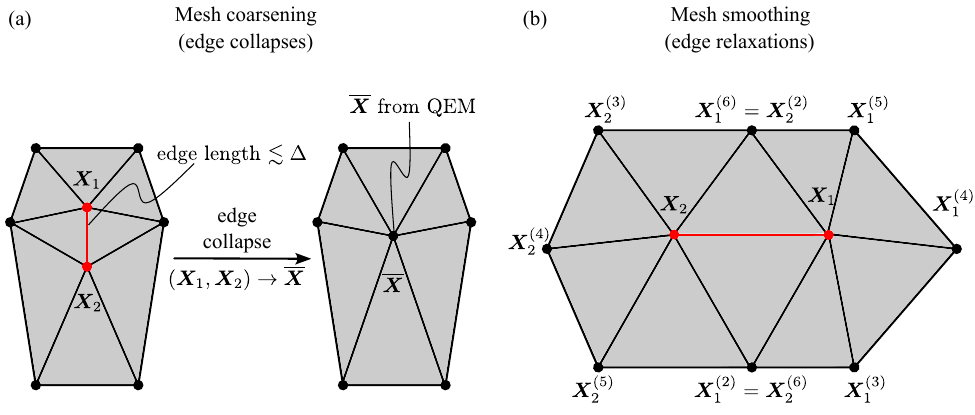}
 	\caption{(a) Illustration of the mesh coarsening procedure. At each timestep, edges are incrementally collapsed based on whether their lengths fall below a threshold of the Eulerian grid spacing ($0.7\Delta$ for our present work). An edge-collapse entails collapsing two vertices, $\bm{X}_1$, $\bm{X}_2$ to a new location $\overline{\bm{X}}$. The location of $\overline{\bm{X}}$ is determined using Quadric Error Metrics (QEM) detailed in \citep{garland1997}. (b) Notation for the mesh smoothing procedure. Here, the edge ($\bm{X}_1,\bm{X}_2$) is flagged for a relaxation, meaning $\bm{X}_1$, $\bm{X}_2$ are free to move to locations computed using information from their vertex neighbours. A volume-conserving correction ensures that the new locations of $\bm{X}_1$, $\bm{X}_2$ do not modify the local volume contributions to the solid object.}
 	\label{fig:fig4_remesh}
\end{figure}
Following the coarsening procedure by edge-collapses in \S \ref{sec:ecol}, it is possible for numerically-undesirable features such as skewed triangles or sharp corners to form on the triangulated geometry. To continue melting the solid in a stable manner, intervention through mesh smoothing is necessary to alleviate this issue. Here, we will adopt the smoothing algorithm for triangulated geometries proposed by \citet{kuprat2001} (specifically, algorithm no. 4 in their work). The main appeal of their algorithm is that it is a volume-conserving formulation, meaning that the triangulated surface will undergo a net-zero change in volume during the smoothing. Such a feature for melting problems is highly desirable, given that the solid volume as a function of time, $V_s(t)$, is typically the primary quantity of interest for prediction. A volume-conserving remeshing procedure will help to ensure that measures of $V_s(t)$ will be uncontaminated by artificial changes to the volume from the remeshing step. As noted by \citet{roghair2016}, the volume-conserving smoothing algorithm of \citet{kuprat2001} can also be recast into a smoothing algorithm that achieves a target volume change. When applied in sequence following the mesh coarsening step (\S \ref{sec:ecol}), one can set the target volume change of the smoothing procedure to be the negative of the volume change incurred during the coarsening step, $\delta V_{\mathrm{smooth}} = -\delta V_{\mathrm{coarsen}}$, such that the net volume change to due to remeshing is zero to machine precision: $\delta V_{\mathrm{coarsen}} + \delta V_{\mathrm{smooth}} = 0$. One can then interpret the smoothing step as also providing a hysteresis restorative effect to the geometry volume. 

The elementary operation of the smoothing algorithm proposed by \citet{kuprat2001} consists of \textit{edge relaxations}. An edge relaxation simply means that its two vertices are free to move during the smoothing procedure. Determining which edges to relax is a free choice determined by the user. Consistent with our philosophy for the coarsening step (\S \ref{sec:ecol}), our approach is to apply minimal changes to the solid geometry, only applying interventional smoothing where necessary. Our present approach has been to flag edges which have been modified by the previous coarsening step as edges which are eligible for relaxations. All other edges remain `anchored' and will not move during smoothing unless the anchored edge shares a vertex with another edge that is flagged for relaxation. The flag for edge relaxations is reset after every timestep and the flagging operation is only triggered if mesh coarsening is applied at the given timestep. This heuristic strategy is based on our observation that skewed triangles and sharp corner features tended to form only in regions that were modified by our mesh-coarsening procedure. Consequently, this smoothing strategy also retains the incremental, local approach that was used for coarsening. 

We now provide details on the computations entailed in the edge relaxations. With reference to figure \ref{fig:fig4_remesh}(b), suppose an edge connected by vertices $\bm{X}_1$, $\bm{X}_2$ is flagged for relaxation. A primitive smoothing algorithm seeks to move these vertices to a new location, $\bm{X}_i^s$, where $i = \{1,2\}$. The simplest smoothing approach is to compute the new smoothed locations, $\bm{X}_i^s$, as the centroid of its vertex neighbours \citep{botsch2010,baerentzen2012}. Following the indexing notation of figure \ref{fig:fig4_remesh}(b):
\refstepcounter{equation}
$$
\bm{X}_1^s = \frac{1}{\mathcal{N}_1}\left( \bm{X}_2^s + \sum_{j=2}^{\mathcal{N}_1} \bm{X}_1^{(j)} \right), \quad \bm{X}_2^s = \frac{1}{\mathcal{N}_2}\left( \bm{X}_1^s + \sum_{j=2}^{\mathcal{N}_2} \bm{X}_2^{(j)} \right) ,
\eqno{(\theequation{a,b})}\label{eq:smooth_xs}
$$
which are two equations to determine the two unknowns $\bm{X}_1^s$, $\bm{X}_2^s$. Here, $\mathcal{N}_1$, $\mathcal{N}_2$ are the number of vertex neighbours for $\bm{X}_1$, $\bm{X}_2$ respectively. This shifts each vertex $i$ by a displacement $\mathrm{d}\bm{X}_i^s = \bm{X}_i^s - \bm{X}_i$. These displacements will cause a change in volume of the geometry in general. The volume-conserving approach corrects for this by an additional incremental shift, $h \mathbf{\hat{e}}$:
\begin{equation}
    \mathrm{d}\bm{X}_i = \mathrm{d}\bm{X}_i^s + h \mathbf{\hat{e}}.
\end{equation}
after which the new locations are $\bm{X}_i^{\mathrm{new}} = \bm{X}_i + \mathrm{d}\bm{X}_i$. The two unknowns $h$, $\mathbf{\hat{e}}$ are determined by two conditions. (1) Enforcing local volume conservation in the neighbourhood of vertices $\bm{X}_1$, $\bm{X}_2$, and (2): minimising $|h \mathbf{\hat{e}}|$ which ensures that the final shift, $\mathrm{d}\bm{X}_i$, is as close as possible to the original shift determined from the primitive smoothing algorithm $\mathrm{d}\bm{X}_i^s$. These conditions lead to explicit expressions for $h$, $\mathbf{\hat{e}}$ \citep{kuprat2001,roghair2016}:
\refstepcounter{equation}
$$
h = \frac{6\delta V \mathrm{sign}(B) - B}{\mathbf{\hat{e}}\bcdot [\bm{A}_1 + \bm{A}_2 + \bm{v} \times (\mathrm{d}\bm{X}_1^2 - \mathrm{d}\bm{X}_2^s)] }, \quad \mathbf{\hat{e}}  = \frac{\bm{A}_1 + \bm{A}_2 + \bm{v} \times (\mathrm{d}\bm{X}_1^s - \mathrm{d}\bm{X}_2^s)}{|\bm{A}_1 + \bm{A}_2 + \bm{v} \times (\mathrm{d}\bm{X}_1^s - \mathrm{d}\bm{X}_2^s)|}
\eqno{(\theequation{a,b})}\label{eq:smooth_he}
$$
where $B$, $\bm{A}_1$, $\bm{A}_2$, $\bm{v}$ are determined using information from the neighbours of $\bm{X}_1$, $\bm{X}_2$. Full details can be found in \citep{kuprat2001}. In (\ref{eq:smooth_he}a), $\delta V$ appears as a corrective desired volume change, which is set as $\delta V = - \delta V_{\mathrm{coarsen}}$ as mentioned above. Note that our expression differs to that of \citet{roghair2016}, who did not include the $\mathrm{sign}(B)$ correction for $h$. The direction of the volume correction is fixed as $\mathbf{\hat{e}}$ with magnitude $h$, with only $h$ explicitly depending on $\delta V$. The direction of this volume correction must appropriately oppose whether the coarsening step has shrunk ($\delta V < 0$) or expanded ($\delta V > 0$) the geometry, and this is accounted for with the sign($B$) correction. When performing multiple edge relaxations in a single step, it is important that the $\delta V$ correction is only applied for a \textit{single} edge relaxation. Once the $\delta V$ correction has been applied, the solid volume has been restored to its original volume prior to remeshing and any further edge relaxations should not alter this volume. Consequently, one should set $\delta V = 0$ in (\ref{eq:smooth_he}a) once the first edge relaxation has been performed at a given timestep.

The algorithm of \citet{kuprat2001} is an iterative one which continually loops over edge relaxations until some specified number of iterations has finished. We have found that no more than roughly 10 iterations are necessary. Any further iterations yielded negligible displacements which were at the sub-Eulerian grid scale, $\lesssim 0.5 \Delta$. Overall, the full remeshing procedure tends to result in negligible changes of the global Lagrangian mesh properties such as the centroid location, surface area, or the computed inertia tensor. This is in part due to the sub-Eulerian displacements of the remeshing but also the local incremental nature of the remeshing, where the vast majority of triangles in the geometry will remain unaltered during a remeshing iteration. This consequently results in a negligible change of the Lagrangian mesh properties.
\section{Numerical solver overview}\label{sec:num_overview}
Having detailed each individual step of the numerical solver in previous sections, we now provide a summary and precise details on the sequencing of all the steps below.
\begin{tcolorbox}
For each RK3 substep $n$
\begin{enumerate}
   \item Tag Eulerian cells as either solid, fluid, or interface cells via ray-tagging.
    \item Solve for intermediate Eulerian field variables $\{ \hat{u},\hat{v},\hat{w},\theta^*\}$ through equations (\ref{eq:goveqn_discrete1}--\ref{eq:goveqn_discrete4}).
    \item IB load evaluation
    \begin{enumerate}
        \item Evaluate direct-forcing $\bm{f}$, $f^\theta$ (\S \ref{eq:mls-ibm}). 
        \item Evaluate surface heat fluxes (\S \ref{sec:stefan}).
        \item Evaluate hydrodynamic loads and torques (\S \ref{sec:fsi}).
    \end{enumerate}
    \item Impose boundary conditions and surface motion
    \begin{enumerate}
        \item Spread forcing: $\bm{u}^* = \bm{\hat{u}} + \alpha_n \Delta t \bm{f},\quad \theta^{n+1} = \theta^* + \alpha_n \Delta t f^\theta$.
        \item Melt the interface via the Stefan condition (\ref{eq:stefan-posupdate}--\ref{eq:stefan-vertex}) using the heat fluxes evaluated at step 3(b).
        \item Apply rotational and translational motion update (\S \ref{sec:fsi}) with the hydrodynamics loads evaluated at step 3(c).
    \end{enumerate}
    \item Apply remeshing (\S \ref{sec:remesh}) if necessary. Recompute Lagrangian mesh properties (triangle areas, normal vectors, edge lengths, etc.).
    \item Pressure projection step, equation \eqref{eq:press_proj}.
\end{enumerate}
\end{tcolorbox}
The overview we have provided coincides with a so-called loosely-coupled approach for solving FSI problems \citep{hou2012}. In a loosely-coupled method, the governing equations for the flow-field and solid-body motion are solved in a sequential manner at each computational timestep. Strongly-coupled approaches by contrast, will iteratively solve the governing equations at each timestep until convergence within a tolerance is reached, entailing a greater computational cost. Typically, loose-coupling is unstable for FSI problems where added-mass effects are important and strong-coupling becomes necessary \citep{hu2001,causin2005,borazjani2008}. Added mass is typically a function of the solid-to-fluid density ratio but also the solid geometry in question \citep{kennard1967,brennen2005}. In our present work, we have found that loose coupling is adequate. This is in part due to the small interface melt velocities and hence relatively slow interface motion. Consequently, the solid-body essentially moves as a rigid body whose dynamics depend only on the resultant hydrodynamic loads.

In our overview, the final operation associated with the Lagrangian mesh is the remeshing step, with no subsequent Lagrangian computations in the remainder of the timestep. This choice helps to minimise the computational overhead associated with the intermediate updates to the Lagrangian mesh geometry occurring between steps $3$--$4$. Following the completion of remeshing, the subsequent updates to the Lagrangian mesh properties (e.g. triangle areas, normal vectors, edge lengths) may then be readily used for the next timestep.
\section{Test cases}\label{sec:tests}
\subsection{Diffusive melting of a fixed sphere}
In this test case, we will consider the diffusive melting of a fixed sphere. This is a simplified problem which does not couple the temperature to flow, i.e. $\bm{u} = \mathbf{0}$ is fixed, nor does it consider FSI. The purpose of this test case is to validate melting of the interface, and also to demonstrate the remeshing feature of the solver. 

For parameters, we consider a sphere of initial diameter $D_0$ immersed in a domain of size $L^3$ at the centre with $D_0 = 0.2L$ and with an initial uniform solid temperature equal to the melt temperature, $\theta_{\mathrm{melt}} = 0.0$ and uniform initial liquid temperature $\theta_l = 1.0$. Fixed Dirichlet boundary conditions $\theta = \theta_l$ are applied at all domain boundaries. The temperature difference $\Delta \theta \equiv \theta_l - \theta_{\mathrm{melt}}$ characterises the speed at which the object melts and is captured in the Stefan number, $St \equiv \mathcal{L} / (c_p \Delta \theta)$ which we fix to be 1.0 for these tests. Three grid resolutions have been tested and are summarized in table \ref{tab:diff}. This includes the Lagrangian mesh resolution which is documented by the initial number of Lagrangian nodes, $N_l$ in the mesh. The ratio of the initial Lagrangian mesh edge lengths to the Eulerian grid spacing are also kept the same for all grid choices. Across all grid sizes, the solid is able to melt down to sufficiently small fractions of the initial solid volume, $V_0$. The final volume is consistently well-below 1\% of the initial solid volume $V_s / V_0 < 0.01$, which for practical purposes related to prediction, is typically more than adequate. The final volume relative to the Eulerian grid cell volume, $\Delta^3$ is also provided in table \ref{tab:diff} where we demonstrate $V_s / \Delta^3 \lesssim 40$. That is, final volumes which approach the scale of the Eulerian grid. Recall that each Lagrangian node spreads its forcing across a local Eulerian support domain of size $(3 \Delta)^3 = 27\Delta^3$ (figure \ref{fig:fig1_formulation}b). Consequently, one expects that the solution accuracy likely deteriorates when the solid volume approaches this size, $V_s / \Delta^3 \approx O(10)$. One may simply choose to terminate the simulation once a threshold $V_s$ has been reached, say, when some threshold fraction of the initial solid volume $V_0$ is reached. 

\begin{figure}
\centering
\includegraphics[]{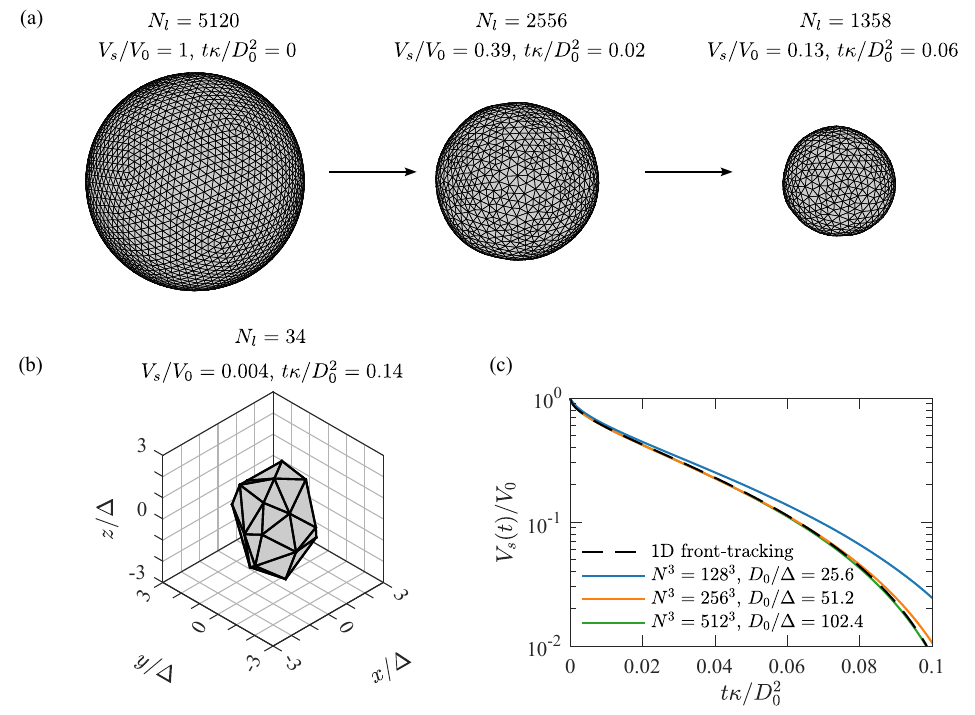}
 	\caption{Results for diffusive melting of a fixed sphere at $St = 1.0$. (a) Solid geometry at various times during melting illustrating the dynamic remeshing procedure for $N^3 = 128^3$. (b) Final solid geometry stage which approaches the scale of Eulerian grid cells. (c) Quantitative validation of the solid volume against time comparing various grid resolutions $N^3$ and results from 1D axisymmetric front-tracking.}
 	\label{fig:fig5_diffusion}
\end{figure}

The key to achieving these small final volumes $V_s / V_0$, lies in the dynamic remeshing procedure which we provide snapshots of in figure \ref{fig:fig5_diffusion}(a). As the solid melts, the geometry is gradually coarsened, reflected in the reduction of the triangle count $N_l$. The coarsening procedure ensures that at each stage, the Lagrangian grid spacing $\Delta_L$ is within a threshold of the Eulerian grid spacing $\Delta$ for numerical stability, maintained all the way down to the final solid geometry stage (figure \ref{fig:fig5_diffusion}b). Quantitative validation of this diffusion problem is presented in figure \ref{fig:fig5_diffusion}(c). Here, we compare results for the solid volume over time from the various grid resolutions to a 1D front-tracking method where details can be found in 
\ref{app:1d_ft}. Excellent agreement is obtained beyond a grid size $N^3 = 256^3$. One key parameter to consider here is the number of Eulerian cells resolving the sphere's initial diameter, $D_0/\Delta$. Typically, $D_0 / \Delta \gtrsim 30$ constitutes a well-resolved simulation for rigid-body simulations \citep{uhlmann2005,breugem2012,kempe2012a}. Absent adaptively refining the Eulerian mesh, maintaining this condition becomes more stringent as the solid inevitably melts. The comparison of figure \ref{fig:fig5_diffusion}(c) demonstrates that so as long as we begin with a well-resolved solid geometry, $D_0 / \Delta \gtrsim 50$ say, one can still achieve adequate accuracy for the solid object's volume throughout the entirety of the melt duration. 
\begin{table}
\centering
\begin{tabular}{c c c c c c c c}
\hline \hline
    & & & \multicolumn{3}{c}{Initial $\Delta_L / \Delta$} & & \\ \cline{4-6} \\ \\[-0.7cm]
     $N^3$ & Initial $N_l$ & $D_0 / \Delta$ & Min.&  Avg. & Max. & Final $V_s(t)/V_0$ & Final $V_s(t)/\Delta^3$ \\ \hline \\[-0.3cm]
     $128^3$ & 5120 & 25.6 & 0.89 & 0.97 & 1.06 & $4.40 \times 10^{-3}$ & 38.6\\
     $256^3$ & 20480 & 51.2 & 0.89 & 0.97 & 1.06 & $1.31 \times 10^{-4}$& 9.2 \\
    $512^3$ & 81920 & 102.4 & 0.89 & 0.97 & 1.06 & $1.85 \times 10^{-7}$ & 0.10 \\ \hline \hline
\end{tabular}
\caption{Parameters and results for diffusive melting of a fixed sphere adopting 3 different discretisations. For all cases, we fix $St = 1.0$, $L = 1.0$, $\kappa = 0.1$. Here, $N^3$ is the Eulerian grid size; initial $N_l$ the initial number of Lagrangian nodes; $D_0/\Delta$ number of Eulerian cells across the initial sphere diameter $D_0$; $\Delta_L / \Delta$ the Lagrangian mesh edge length normalised on the Eulerian grid spacing. The final solid volume $V_s$ normalised on the initial volume $V_0$ and Eulerian cell volume $\Delta^3$ is also provided. All cases adopt a fixed timestep $\Delta t = 10^{-5}$.}
\label{tab:diff}
\end{table}
\subsection{Neutrally-buoyant melting of a fixed sphere}
\begin{figure}
\centering
		\includegraphics[]{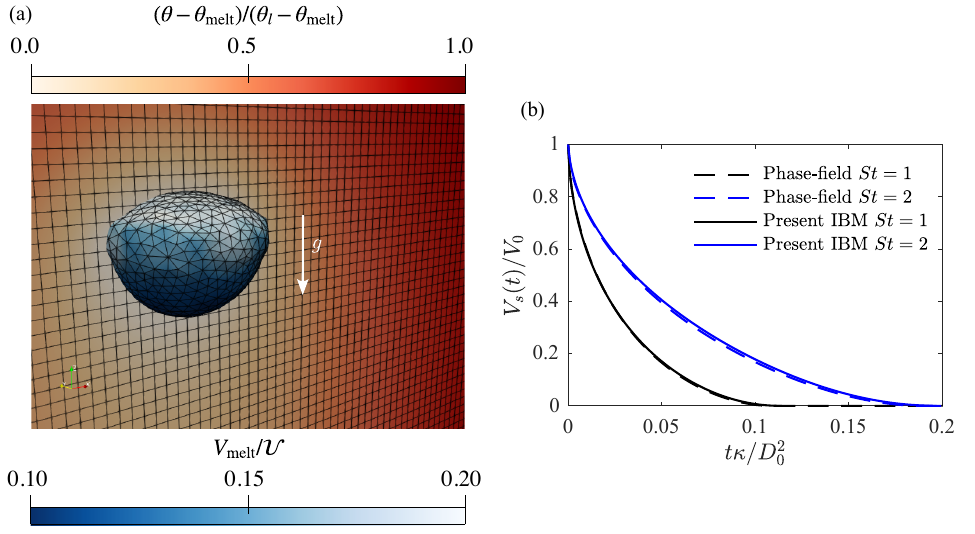}
 	\caption{(a) Instantaneous snapshot showing neutrally-buoyant melting of a fixed sphere at $Ra = 10^3$, $St = 1.0$. The dimensionless temperature field at the $y = L/2$ midplane is shown as well as the local melt velocity variation made dimensionless with the freefall velocity $\mathcal{U} \equiv \sqrt{\beta g D_0 \Delta \theta}$. (b) Quantitative comparison of the solid volume versus time for $St = 1.0$, $2.0$ for a fixed $Ra = 10^3$, comparing the present approach to phase-field simulations using the scheme of \citet{hester2020}.}
 	\label{fig:fig6_buoy}
\end{figure}
We consider the neutrally-buoyant melting of a fixed sphere. In this problem, FSI is not considered (i.e. the solid does not translate or rotate), but melting is now coupled to Navier--Stokes. For neutrally-buoyant melting conditions, the density of the solid and liquid remain matched, but thermal buoyancy is permitted. This is introduced through an additional body force in the momentum equation \eqref{eq:navstokes} in the form of the Boussinesq approximation, $\bm{\mathcal{F}} = \beta g  \Delta \theta \mathbf{\hat{e}}_z$ where $\beta$ is the thermal expansion coefficient of the liquid, $g$ is gravity, and $\mathbf{\hat{e}}_z$ is the unit vector in the $z$-direction in which gravity acts. The dimensionless numbers governing this problem alongside the Stefan number $St$, are the Rayleigh number, $Ra$, and Prandtl number, $Pr$:
\refstepcounter{equation}
$$
Ra \equiv \dfrac{\beta g \Delta \theta D_0^3}{\nu \kappa}, \qquad Pr \equiv \frac{\nu}{\kappa}.
\eqno{(\theequation{a,b})}\label{eq:ra_pr}
$$
We consider a sphere of initial diameter $D_0$ immersed in the centre of the domain of size $L^3$. We set $D_0 = 0.1L$ and the initial temperature inside the solid to be the melt temperature, $\theta_{\mathrm{melt}}$. The Rayleigh number and Prandtl number are fixed as $Ra = 10^3$, $Pr = 1.0$ and we have considered two different choices of $St = 1.0$ and $St = 2.0$, corresponding to relatively faster and slower melting respectively. The timestep is fixed as $\Delta t = 10^{-3}$ corresponding to $\Delta t / \mathcal{T} = 3.2 \times 10^{-3}$ where $\mathcal{T} \equiv D_0 / \mathcal{U} \equiv \sqrt{D_0 / (\beta g \Delta \theta)}$ is the freefall timescale. The Eulerian grid size is fixed as $512^3$ for the domain size $L = 1.0$, which has $D_0 / \Delta =  51.2$ Eulerian cells across the initial sphere diameter. A snapshot of the flow for $St = 1.0$ is shown in figure \ref{fig:fig6_buoy}(a).
The fluid around the solid is cooled, leading it to flow downwards around the solid body.
The accumulation of cool fluid on the underside of the solid insulates this area from the warm ambient fluid, so that it melts more slowly than the upper side and induces top-down asymmetry.
One sees that this intuition is corroborated by the local interface melt velocity magnitude $V_{\mathrm{melt}} \equiv -\bm{U}_{\mathrm{melt}} \bcdot \bnhat$ being smaller around the bottom of the solid object. Quantitative comparisons for this problem are provided in figure \ref{fig:fig6_buoy}(b). Here, we make comparisons of our present method to results from solving this problem adopting the phase-field implementation of \citet{hester2020}. Three of the present authors have adopted this phase-field approach in prior work on melting \citep{yang2023a,yang2023b,yang2024} and the approach has been validated \citep{howland2024}. Across both cases, the Eulerian grid resolution has been matched, namely, $N^3 = 512^3$. We observe excellent agreement across both $St = 1.0$, $St = 2.0$ for the full melt duration.
\subsection{Lagrangian melting in homogeneous isotropic turbulence}
\begin{figure}
\centering
		\includegraphics[]{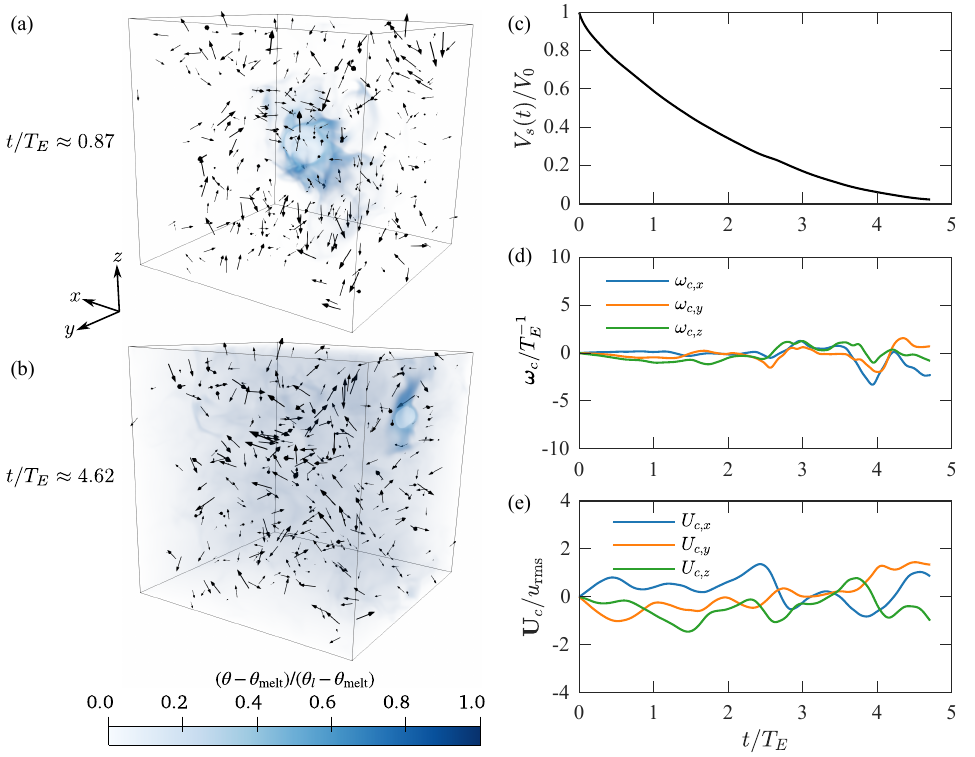}
 	\caption{Lagrangian melting in homogeneous isotropic turbulence (HIT) at Taylor Reynolds number $Re_\lambda \approx 100$, $St = 1.0$, $Pr = 1.0$. (a,b) Volume contours of temperature in the domain visualised with instantaneous velocity vectors at different instances during the melt duration. The colour for the background liquid temperature $\theta = \theta_l$ is made transparent for visibility. (c) Solid object volume versus time. (d) Centroid angular velocity; (e) centroid translational velocity. For (c--e), normalisations are done on the r.m.s. velocity $u_{\mathrm{rms}}$ and large-eddy timescale, $T_E \equiv u_{\mathrm{rms}}^2/\varepsilon$ where $\varepsilon$ is the turbulent dissipation rate.}
 	\label{fig:fig7_hit}
\end{figure}
We now consider the full Lagrangian melting problem, solving the full set of equations outlined in \S \ref{sec:goveqn}. We simulate the melting of a freely-moving sphere in homogeneous isotropic turbulence (HIT), as visualised in figure \ref{fig:fig7_hit}(a,b). The HIT forcing employed is the stochastic forcing scheme of \citet{eswaran1988} which is a stable choice for interface-resolved, particle-laden HIT simulations \citep{chouippe2015}. Temperature is treated as a passive scalar with $Pr = 1.0$ and we fix $St = 1.0$. The initial sphere radius is $R_0 = 0.1L$ and a grid size of $N^3 = 256^3$ is employed. We use a fixed timestep $\Delta t = 5 \times 10^{-4}$ corresponding to $\mathrm{CFL} \lesssim 0.15$. To initialise the problem, we fix the sphere at the centre of the domain and treat it as a rigid-body (i.e. no melting) until a statistically stationary flow has developed. Once this state is achieved, solid-body rotation, translation and melting of the sphere are enabled, while the HIT forcing is maintained. The average Taylor Reynolds number through the duration of melting is $Re_\lambda \approx 100$.

In figure \ref{fig:fig7_hit}(c,d,e), we present time traces of the solid object's volume, angular velocity, and centroid velocity across the melt duration. The solid object undergoes stochastic rotations and translations to be expected from a forced turbulent flow. In tandem with this continual motion, the solid object successfully melts down to a final volume that is a negligible fraction of the initial volume $V_s / V_0 \lesssim 0.02$ (figure \ref{fig:fig7_hit}c), thus demonstrating our present method as being suitable for Lagrangian melting.
\section{Conclusions}\label{sec:conclusions}
We have presented a front-tracking IB framework, suitable for simulating Lagrangian melting problems. Our method adopts the MLS-IB approach, which has been shown to be capable of handling a diverse range of FSI problems \citep{vanella2009,detullio2016,wang2019,viola2020}. The novelty of our present work has been demonstrating first, how the MLS framework can be extended to incorporate the Stefan melting condition. Second, we have demonstrated that the key to enabling these Lagrangian melting simulations lies in a dynamic remeshing procedure which serves to maintain a consistent Lagrangian-to-Eulerian grid and spacing ratio, ensuring stable interpolation and spreading operations can be maintained between the Eulerian and Lagrangian grid throughout the melting duration. Our remeshing procedure is built upon local, incremental edge-collapse and edge-relaxation operations, making it a scalable approach as these operations typically comprise a negligible computation burden per timestep. As an additional advantageous feature, our remeshing procedure is one which can ensure volume-conservation (i.e. no net volume-change during the remeshing step) following prior works \citep{kuprat2001,roghair2016}, which, for melting problems especially is a highly-attractive feature given that the solid volume as a function of time is the primary quantity of interest for prediction.

An immediate point arising in extending our present work towards more general melting problems concerns the treatment of variable densities and thermal conductivities between the solid and liquid phases. Such conditions would be relevant to ice-water systems for instance, where $\rho_{\mathrm{ice}}/\rho_{\mathrm{water}} \approx 0.9$, $k_{\mathrm{ice}}/k_{\mathrm{water}} \approx 4.0$ at room temperature and $k$ denotes the thermal conductivity. Although variable material properties substantially complicate our present method and is why we have left such considerations to future work, we remark that a rich body of literature exists in the numerical simulation of conjugate heat transfer problems \citep{verzicco2002,orlandi2016,mirjalili2022} and phase-change problems with variable material properties \citep{esmaeeli2004a,esmaeeli2004b,lyu2021,roccon2024}. Such work may form a blueprint in how our present method might be generalised to handle variable material properties for melting, and we provide a proof-of-concept outline in \ref{app:variable-density-cond} for the primary changes necessary in the governing equations. Regarding the volume-conserving remeshing procedure of \citet{kuprat2001} which constitutes one of the main features of our present method, we anticipate that this could be used without issue when generalised for variable material properties. The algorithm of \citet{kuprat2001} is purely geometrically-driven, only seeking to conserve the volume bounded by the Lagrangian mesh: a computation which is independent of the solid and fluid density. More delicate consideration is required when one wishes to enforce mass conservation for variable-density phase-change flows however, as has been described in existing work \citep{juric1998,esmaeeli2004a,esmaeeli2004b,lyu2021,roccon2024} and which we also describe in \ref{app:variable-density-cond}.
\section*{CRediT authorship contribution statement}
\textbf{Kevin Zhong}: Writing -- original draft, conceptualization, methodology, software, validation, formal analysis, investigation, visualization. \textbf{Christopher J. Howland}: Conceptualization, methodology, formal analysis, writing -- review \& editing. \textbf{Detlef Lohse}: Conceptualization, resources, writing -- review \& editing, supervision, project administration, funding acquisition. \textbf{Roberto Verzicco}: Conceptualization, resources, writing -- review \& editing, supervision.
\section*{Declaration of competing interest}
The authors declare that they have no known competing financial interests or personal relationships that could have appeared to influence the work reported in this paper.
\section*{Data availability}
Data can be provided upon request.
\section*{Acknowledgements}
This work is funded by the European Union (ERC, MultiMelt, 101094492). This work was carried out on the Dutch national e-infrastructure with the support of SURF Cooperative. We acknowledge the EuroHPC Joint Undertaking for awarding the project EHPC-REG-2023R03-178 access to the EuroHPC supercomputer Discoverer, hosted by Sofia Tech Park (Bulgaria). K. Zhong acknowledges Ms. G. Piumini and Dr. N. Hori for the helpful discussions which improved this work. K. Zhong also acknowledges Dr. D. Xu for providing phase-field data for comparative validation.
\appendix

\section{1D front-tracking for spherically symmetric diffusive melting}\label{app:1d_ft}
For diffusive-melting of a spherical solid, the problem is radially-symmetric and can be reduced to a 1D-problem for the radial temperature distribution $\theta(r,t)$ and interface location $R(t)$. The governing equation is
\begin{equation}\label{eq:1d_stefan1}
    \frac{\partial \theta}{\partial t} = \kappa \left( \frac{\partial^2 \theta}{\partial r^2} + \frac{2}{r}\frac{\partial \theta}{\partial r}\right),
\end{equation}
and the boundary conditions to satisfy at the interface are
\refstepcounter{equation}
$$
\theta(r=R) = \theta_{\mathrm{melt}}, \qquad \frac{\mathcal{L}}{c_p}\frac{\mathrm{d}R}{\mathrm{d}t} = \kappa \frac{\partial \theta}{\partial r} \bigg |_{\mathrm{solid}} - \kappa \frac{\partial \theta}{\partial r} \bigg |_{\mathrm{liquid}}.
\eqno{(\theequation{a,b})}\label{eq:1d_rad_stefan}
$$
The domain considered is a finite domain $r \in [0,L]$. The boundary conditions for the domain are:
\refstepcounter{equation}
$$ 
\frac{\partial \theta}{\partial r} \bigg |_{r=0} = 0, \quad \theta(r = L) = \theta_l.
\eqno{(\theequation{a,b})}\label{eq:1d_stefan_bc}
$$
Here, (\ref{eq:1d_stefan_bc}a) imposes a symmetry condition at the radial origin $r=0$ while the Dirichlet condition (\ref{eq:1d_stefan_bc}b) fixes the ambient liquid temperature $\theta_l$. We solve equations (\ref{eq:1d_stefan1}--\ref{eq:1d_stefan_bc}) numerically using a simple 1D front-tracking finite difference method using a large number of grid points, $N = 2048$. This is compared against the full 3D solution of our present method for a spherical solid geometry in figure \ref{fig:fig5_diffusion}(c) where we solve equation \eqref{eq:goveqn2} (with $\bm{u} = \mathbf{0}$) with boundary conditions (\ref{eq:stefan_condition}--\ref{eqn:bc_tmelt}) which are the analogous 3D boundary conditions to (\ref{eq:1d_rad_stefan}a,b). 
\section{Variable conductivity and density treatment}\label{app:variable-density-cond}
In this appendix, we will sketch the primary changes that would be necessary when one wishes to consider differing conductivities and densities between the liquid and solid phase. Broadly, we can categorise the necessary steps into two separate treatments. First, the consideration of variable material properties (e.g. density, conductivity) in solving for the flow field, and second, the treatment of phase-change when variable material properties are considered. For the first consideration, the equations governing heat transport in (\ref{eq:goveqn2}--\ref{eq:goveqn_tsolid}) are replaced by the counterparts:
    \begin{align}
        \rho_l c_{p,l}\left[ \dfrac{\partial \theta}{\partial t} + \bnabla \bcdot (\bm{u}\theta)\right] &= \bnabla \bcdot (k_l \bnabla \theta) + f^\theta  \quad \text{in the liquid phase,} \label{eq:var-dc1} \\ 
        \rho_s c_{p,s}\left[ \dfrac{\partial \theta}{\partial t} + \bnabla \bcdot (\bm{U}_s\theta)\right] &= \bnabla \bcdot (k_s \bnabla \theta) + f^\theta  \quad \text{in the solid phase}, \label{eq:var-dc2}
    \end{align}
    where the $l$ and $s$ subscripts are introduced to denote liquid and solid values respectively. In one-fluid formulations, we are required to compute the material properties by way of an indicator function, $I(\bm{x},t)$, which identifies regions as either solid or fluid for use in (\ref{eq:var-dc1}--\ref{eq:var-dc2}). Taking $I = 1$ in the liquid and $I=0$ in the solid respectively, the local density for instance may be computed as an arithmetic average:
    \begin{equation}
        \rho(\bm{x},t) = I\rho_l + (1-I)\rho_s
    \end{equation}
    and similarly for the specific heat capacities, $c_{p,l}$, $c_{p,s}$, as well as thermal conductivities $k_l$, $k_s$. Although alternatively, the use of harmonic averages for the thermal conductivity has also been demonstrated as a robust option \citep{ferziger2003,tryggvason2011,deising2016}. In front-tracking methods, the Eulerian indicator field $I(\bm{x},t)$ for each computational cell must be computed based on information from the Lagrangian mesh. This can be done using a ray-tagging procedure for instance \citep{orourke1998}, as has been done in the present work. Another popular choice in the front-tracking literature has been to compute $I(\bm{x},t)$ as the solution to a Poisson equation \citep{juric1998,tryggvason2001,shin2002,tryggvason2011}. With knowledge of the material property fields $\left\{ \rho, c_p, k\right\}$ established, equations (\ref{eq:var-dc1}--\ref{eq:var-dc2}) may then be integrated in time to the next timestep.

    For changes related to the equations governing the Lagrangian motion, the effects of variable density are easily incorporated through the addition of a buoyancy/Archimedes force to the right-hand side of \eqref{eqn:ne1}: $(\rho_s - \rho_l)V_s \mathbf{g}$, as is standard in multiphase flow formulations \citep{breugem2012,ardekani2016}.
    
    Next we consider phase-change. For variable density and conductivities, the Stefan condition governing the motion of the melting interface in \eqref{eq:stefan_condition} is replaced by its more general form \citep{carslaw1959,ozicsik1980,worster2000,wells2011}:
    \begin{equation}
    \mathcal{L}\rho_s \bm{U}_{\mathrm{melt}}  = \left( k_s \frac{\partial \theta}{\partial n} \bigg |_{\mathrm{solid}} - k_l \frac{\partial \theta}{\partial n} \bigg |_{\mathrm{liquid}}\right) \mathbf{\hat{n}}.
    \end{equation}
    An additional change is required at the melting interface, and consequently, the velocity boundary condition to be imposed. Owing to the density contrast, $\rho_s \neq \rho_l$, freshly-melted solid at the interface is displaced by the liquid phase. When one considers mass conservation at this melting interface \citep{juric1998}, one can relate the liquid velocity adjacent to the interface, $\bm{u}\bcdot \mathbf{\hat{n}}$ to the interface melt velocity $\bm{U}_{\mathrm{melt}}$ and kinematic interface velocity $\bm{U} \equiv \bm{U}_c + \bm{\Omega}_c \times \bm{r}$:
    \begin{align}\label{eq:var-dc3}
        &\rho_l(\bm{u} - \bm{U}_{\mathrm{melt}})\bcdot \mathbf{\hat{n}} = \rho_s (\bm{U} - \bm{U}_{\mathrm{melt}}) \bcdot \mathbf{\hat{n}} \nonumber \\
        \implies& \bm{u} \bcdot \mathbf{\hat{n}} = \frac{\rho_s}{\rho_l}\bm{U} \bcdot \mathbf{\hat{n}} + \left( 1 - \frac{\rho_s}{\rho_l}\right) \bm{U}_\mathrm{melt} \bcdot \mathbf{\hat{n}}. 
    \end{align}
    which can be incorporated as a velocity boundary condition for the local interface-normal velocity like in \eqref{eq:goveqn_bc_u1}. Note that for $\rho_l = \rho_s$, equation \eqref{eq:var-dc3} correctly reduces to the normal projection of no-slip like in \eqref{eq:goveqn_bc_u1}. Finally, although the flow remains incompressible in the liquid phase, this will no longer hold locally at the interface owing to the liquid displacement effects embodied in \eqref{eq:var-dc3} \citep{esmaeeli2004a}. Consequently, $\bnabla \bcdot \bm{u} \neq 0$ at the interface, which necessitates modifications to the usual solenoidal projection step of incompressible flows. The numerical treatment of this modification has been described by various authors in the context of phase-change flows \citep{juric1998,esmaeeli2004a,esmaeeli2004b,lyu2021,roccon2024}.

\bibliographystyle{elsarticle-num-names}

\bibliography{references}





\end{document}